\newif\ifemulateapj
\emulateapjtrue

\ifemulateapj


\documentclass[iop]{emulateapj}

\newcommand{\eqnewline}[1]{\nonumber \\ #1}

\else
\documentclass[12pt,preprint]{aastex}
\newcommand{\eqnewline}[1]{}

\fi

\newcommand{\Alfven}{Alfv\'en }
\renewcommand{\div}{\mathbf{\nabla} \cdot}
\newcommand{\rot}{\mathbf{\nabla}\times}
\newcommand{\rmR}{\rm R}
\newcommand{\rmL}{\rm L}

\newcommand{\FigureOne}{
\begin{figure}[tb]
 \figurenum{1}
 \ifemulateapj
 \plotone{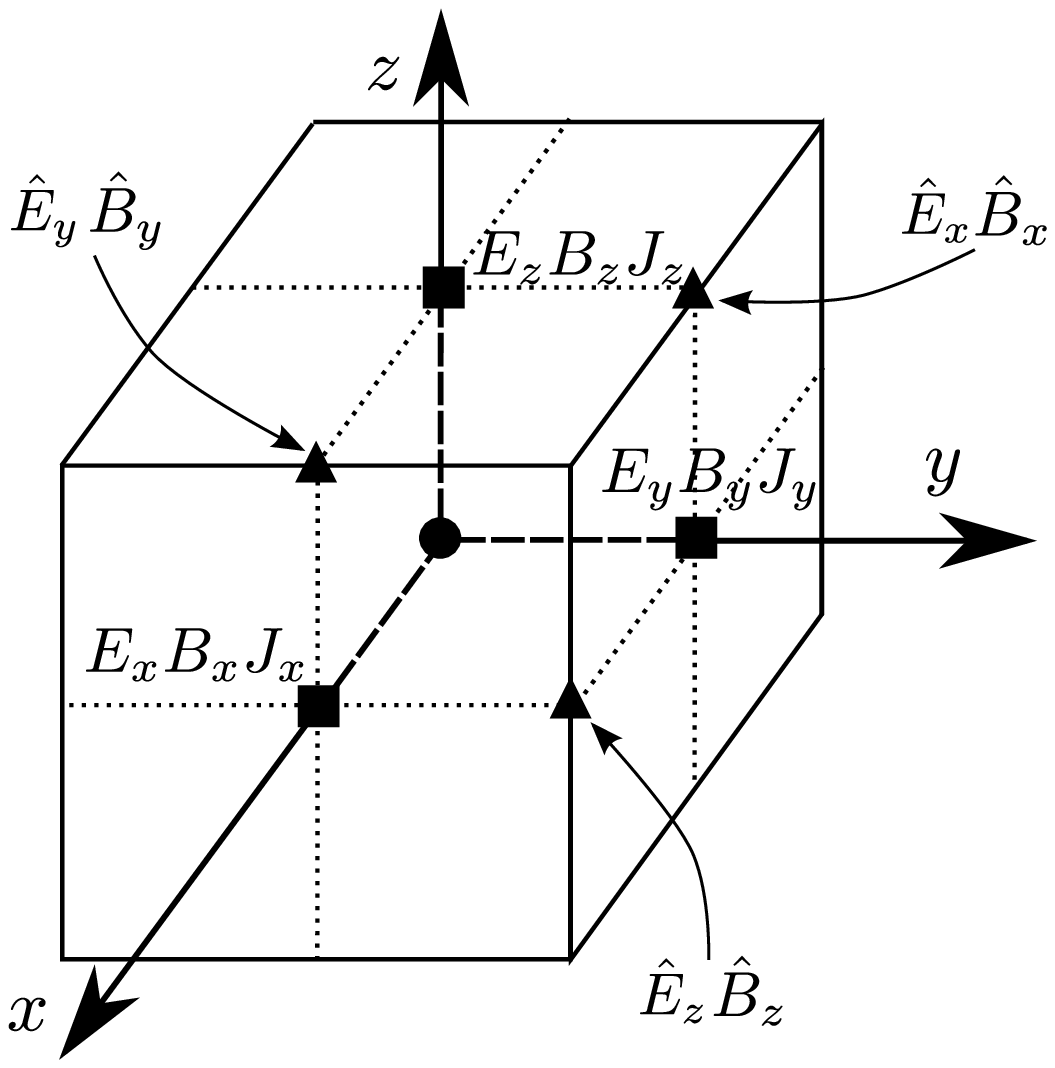}
 \else
 \begin{center}
  \includegraphics[scale=0.75]{figure/fig01.eps}
 \end{center}
 \fi

 \caption{Collocation of physical quantities on a mesh. The two-fluid
 quantities are defined at the point of the filled circle. The primary
 electromagnetic field is defined on the face center (filled squares) in the
 normal direction for each component. The numerical flux for Maxwell's
 equations (denoted by the hat) are defined at the edge center (fllled triangles).}

 \label{fig:collocation}
\end{figure}
}

\newcommand{\FigureTwo}{
\begin{figure*}[tb]
 \figurenum{2}
 \ifemulateapj
 \plotone{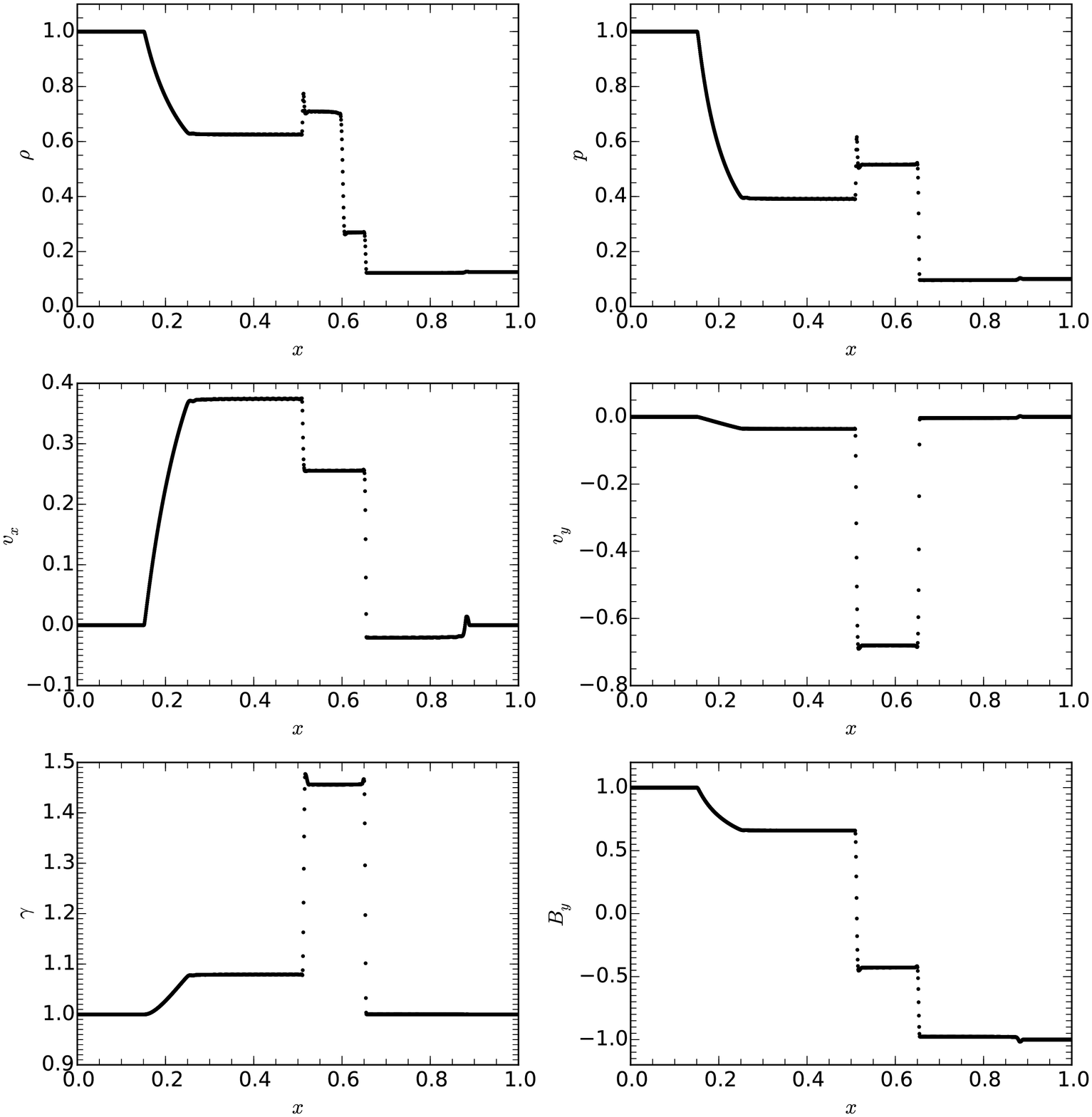}
 \else
 \begin{center}
  \includegraphics[scale=0.60]{figure/fig02.eps}
 \end{center}
 \fi

 \caption{Result for generalized Brio-Wu shock tube problem in RMHD
 regime. The fluid quantities are appropriate averages of the two fluids for
 comparison with published RMHD results.}

 \label{fig:bw_mhd}
\end{figure*}
}

\newcommand{\FigureThree}{
\begin{figure}[tb]
 \figurenum{3}
 \ifemulateapj
 \plotone{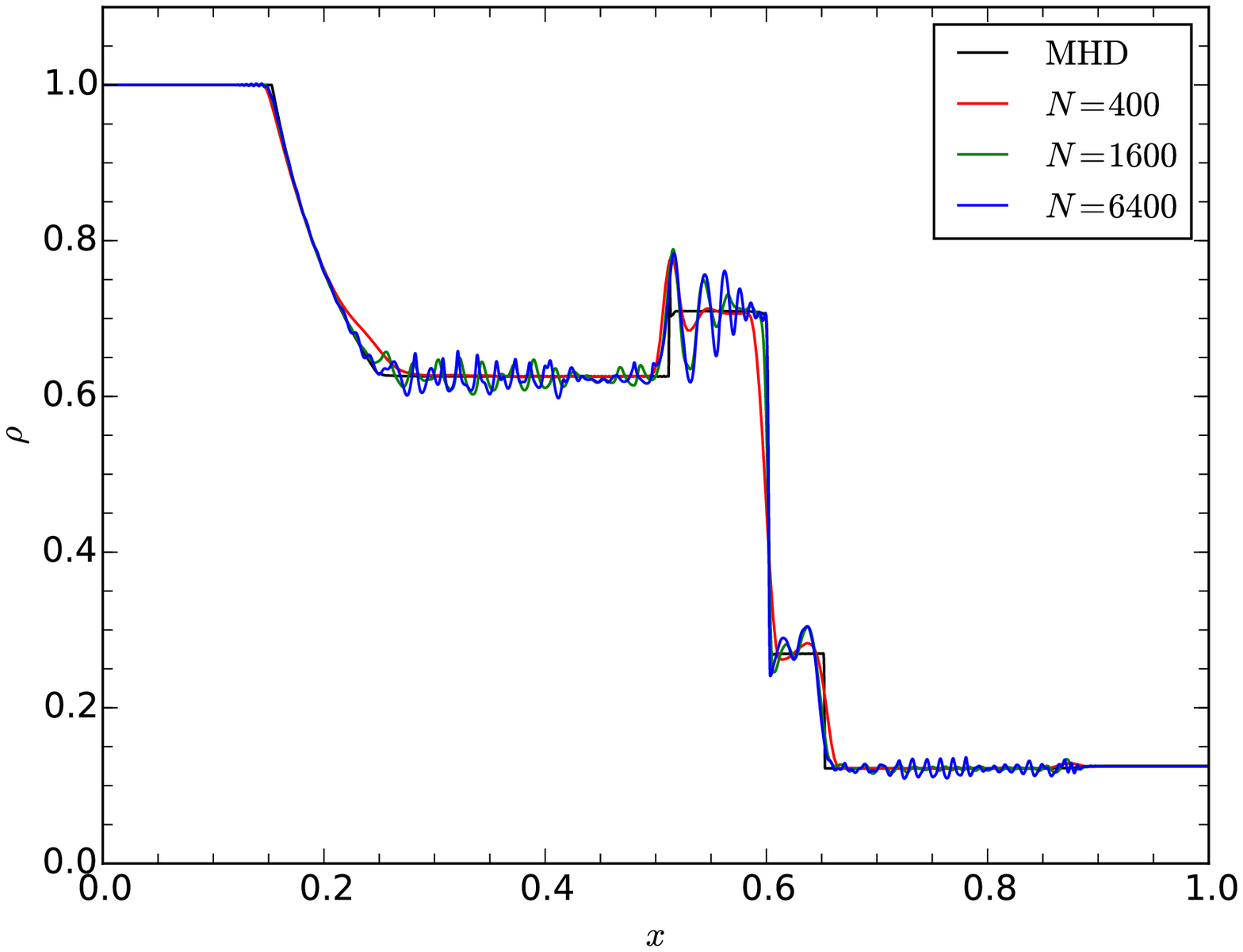}
 \else
 \begin{center}
  \includegraphics[scale=0.75]{figure/fig03.eps}
 \end{center}
 \fi

 \caption{Comparison of total density profiles at $t = 0.4$ obtained with
 different resolutions for generalized Brio-Wu shock tube problem. A pair
 plasma is assumed. Three different numerical solutions with different numbers
 of grid points are shown, along with a reference solution corresponding to the
 RMHD regime.}

 \label{fig:bw_pair}
\end{figure}
}

\newcommand{\FigureFour}{
\begin{figure}[tb]
 \figurenum{4}
 \ifemulateapj
 \plotone{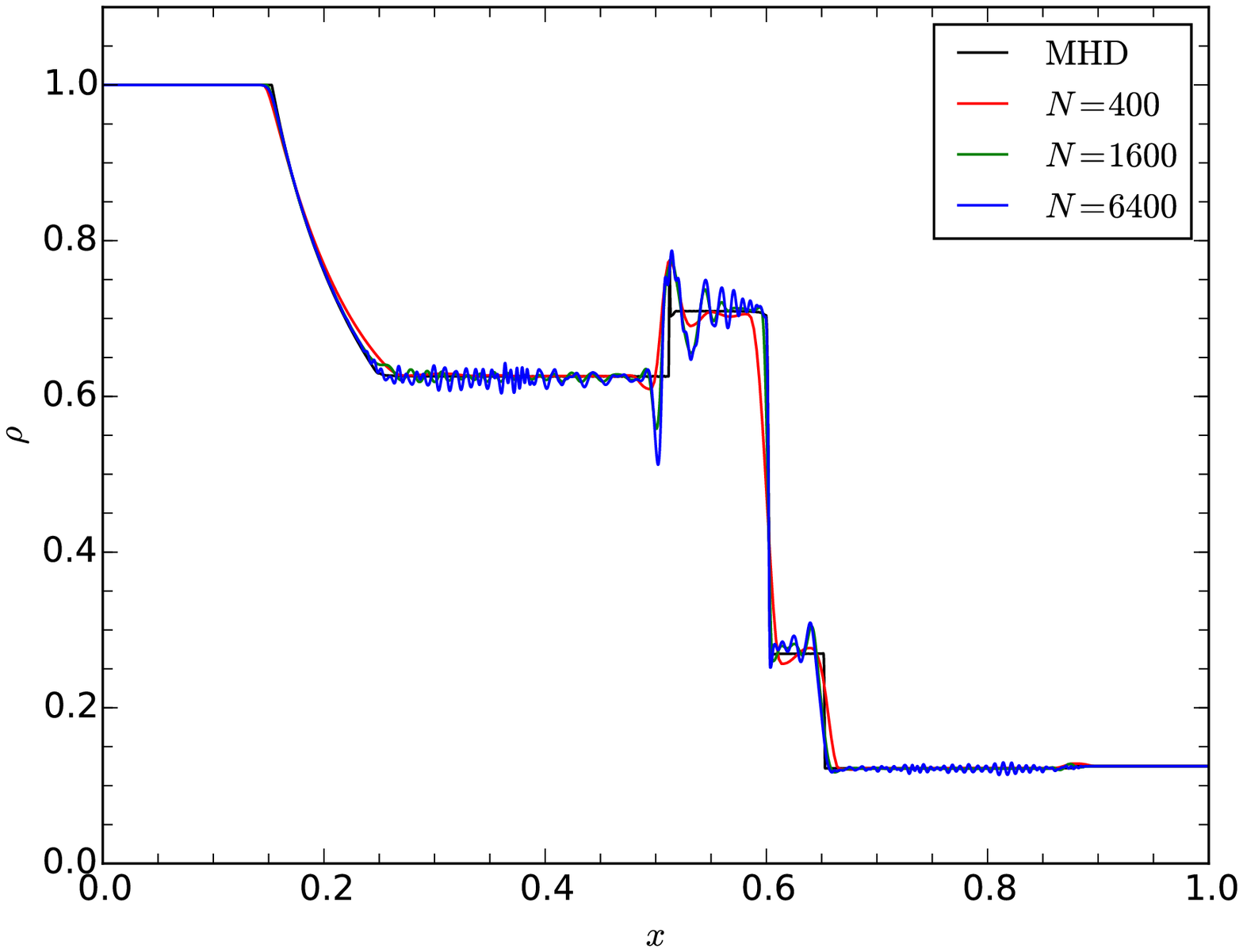}
 \else
 \begin{center}
  \includegraphics[scale=0.75]{figure/fig04.eps}
 \end{center}
 \fi

 \caption{Same as Fig.~\ref{fig:bw_pair} but for an electron-proton plasma
 $m_p/m_e = 100$.}

 \label{fig:bw_ep}
\end{figure}
}

\newcommand{\FigureFive}{
\begin{figure*}[tb]
 \figurenum{5}
 \ifemulateapj
 \plotone{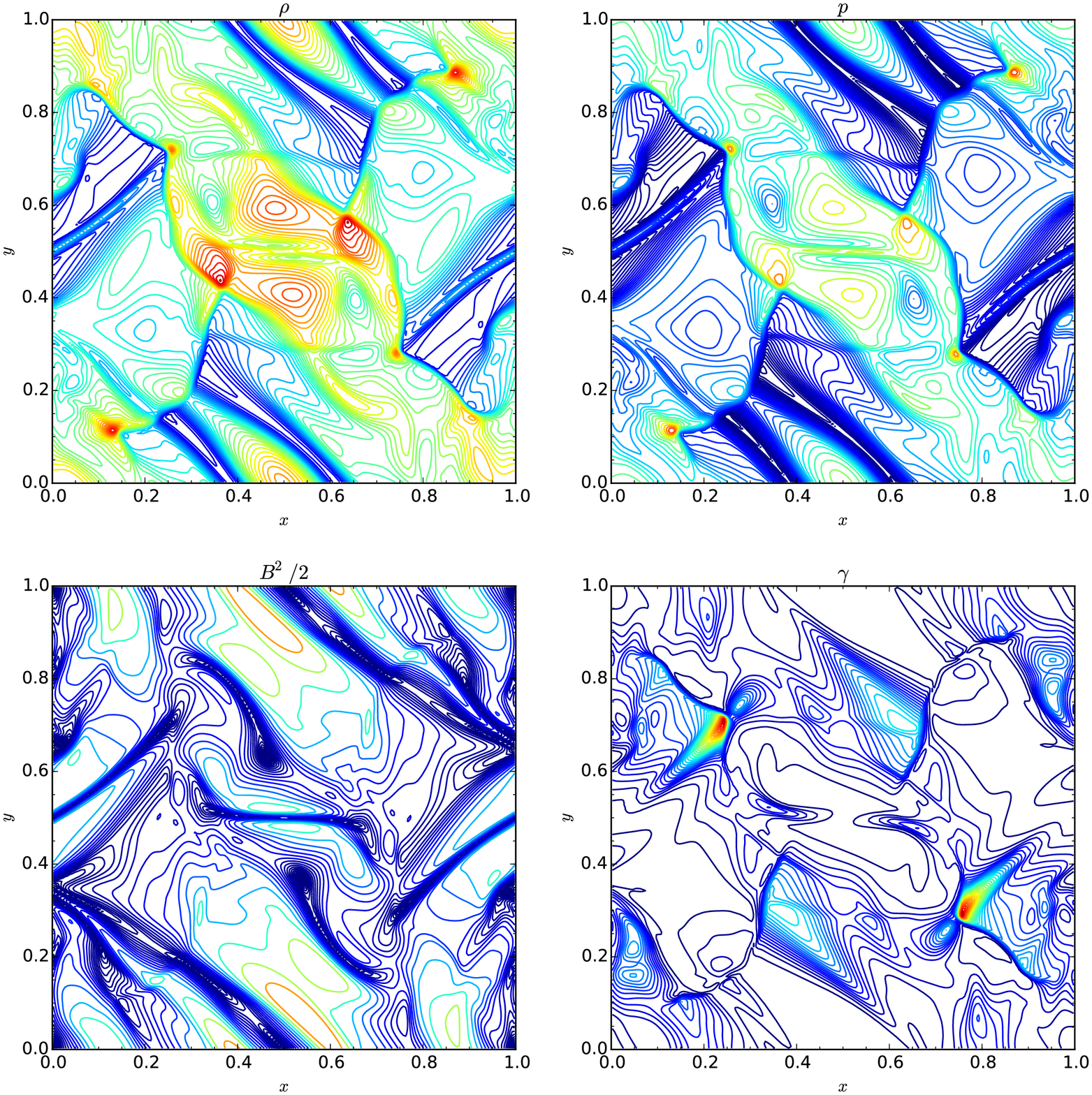}
 \else
 \begin{center}
  \includegraphics[scale=0.40]{figure/fig05.eps}
 \end{center}
 \fi

 \caption{Results for Orszag-Tang vortex problem at $t = 1.0$. The total
 density (top left), gas pressure (top right), magnetic pressure (bottom
 left), and bulk Lorentz factor (bottom right) are shown. In each panel, 40
 contours are drawn in the ranges $5.0 \times 10^{-2}$ to $5.0 \times
 10^{-1}$ for the density, $2.0 \times 10^{-2}$ to $6.0 \times 10^{-1}$ for
 the gas pressure, $3.0 \times 10^{-6}$ to $3.0 \times 10^{-1}$ for the
 magnetic pressure, and $1.01$ to $2.0$ for the Lorentz factor,
 respectively. The contours are evenly spaced in linear scale for the density
 and Lorentz factor, and in logarithmic scale for the gas pressure and
 magnetic pressure.}

 \label{fig:ot}
\end{figure*}
}

\newcommand{\FigureSix}{
\begin{figure*}[tb]
 \figurenum{6}
 \ifemulateapj
 \plotone{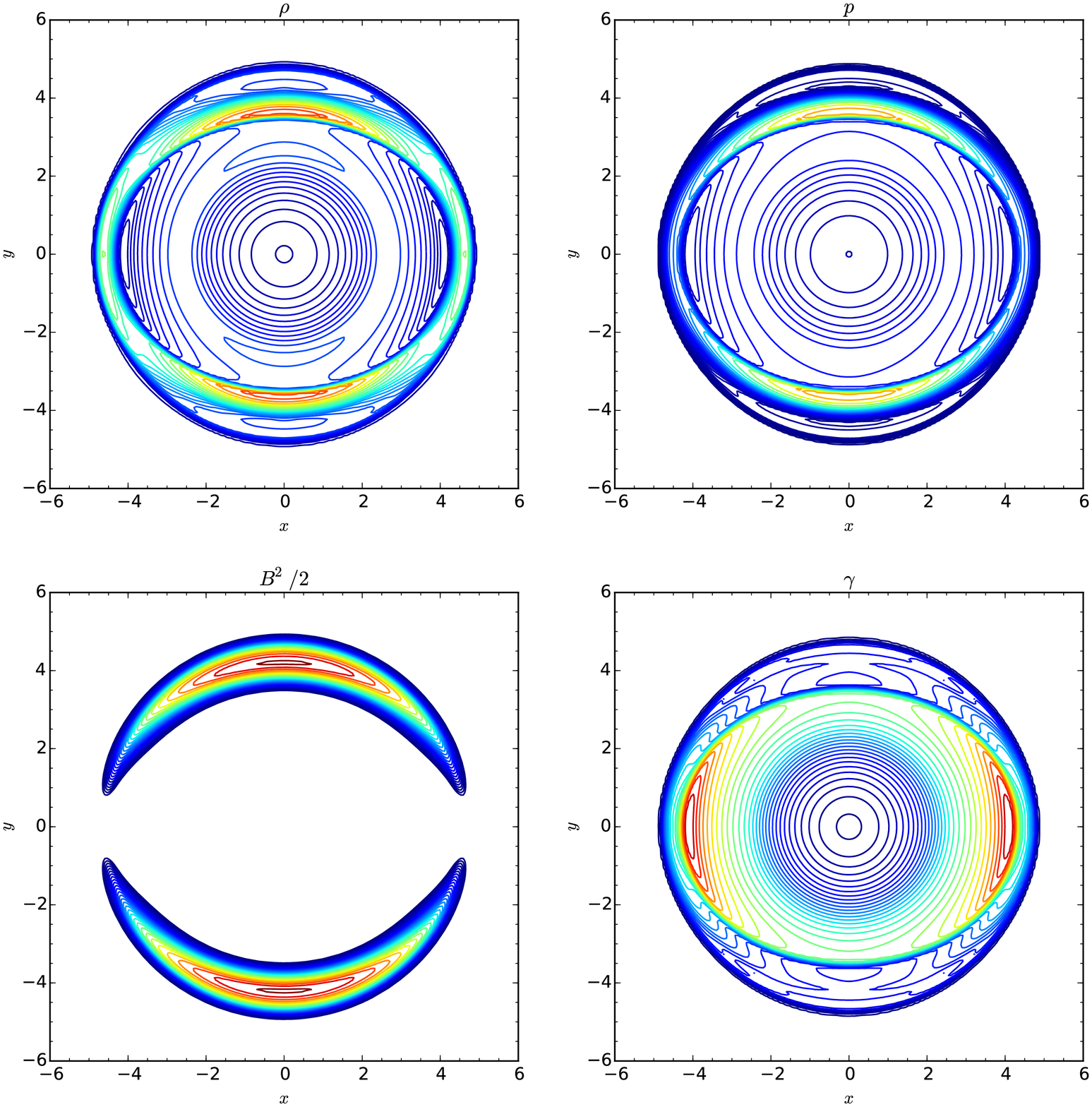}
 \else
 \begin{center}
  \includegraphics[scale=0.40]{figure/fig06.eps}
 \end{center}
 \fi

 \caption{Results for strong cylindrical explosion at $t = 4.0$ in a weakly
 magnetized medium ($B_0 = 0.1$). The format is the same as Figure 5. In each
 panel, 30 contours are drawn in the ranges $8.0 \times 10^{-5}$ to $8.0
 \times 10^{-4}$ for the density, $4.0 \times 10^{-4}$ to $4.0 \times 10^{-2}$
 for the gas pressure, $7.0 \times 10^{-3}$ to $7.0 \times 10^{-2}$ for the
 magnetic pressure, and $1.01$ to $4.0$ for the Lorentz factor,
 respectively. The contours are evenly spaced in logarithmic scale.}

 \label{fig:blast_weak}
\end{figure*}
}

\newcommand{\FigureSeven}{
\begin{figure*}[tb]
 \figurenum{7}
 \ifemulateapj
 \plotone{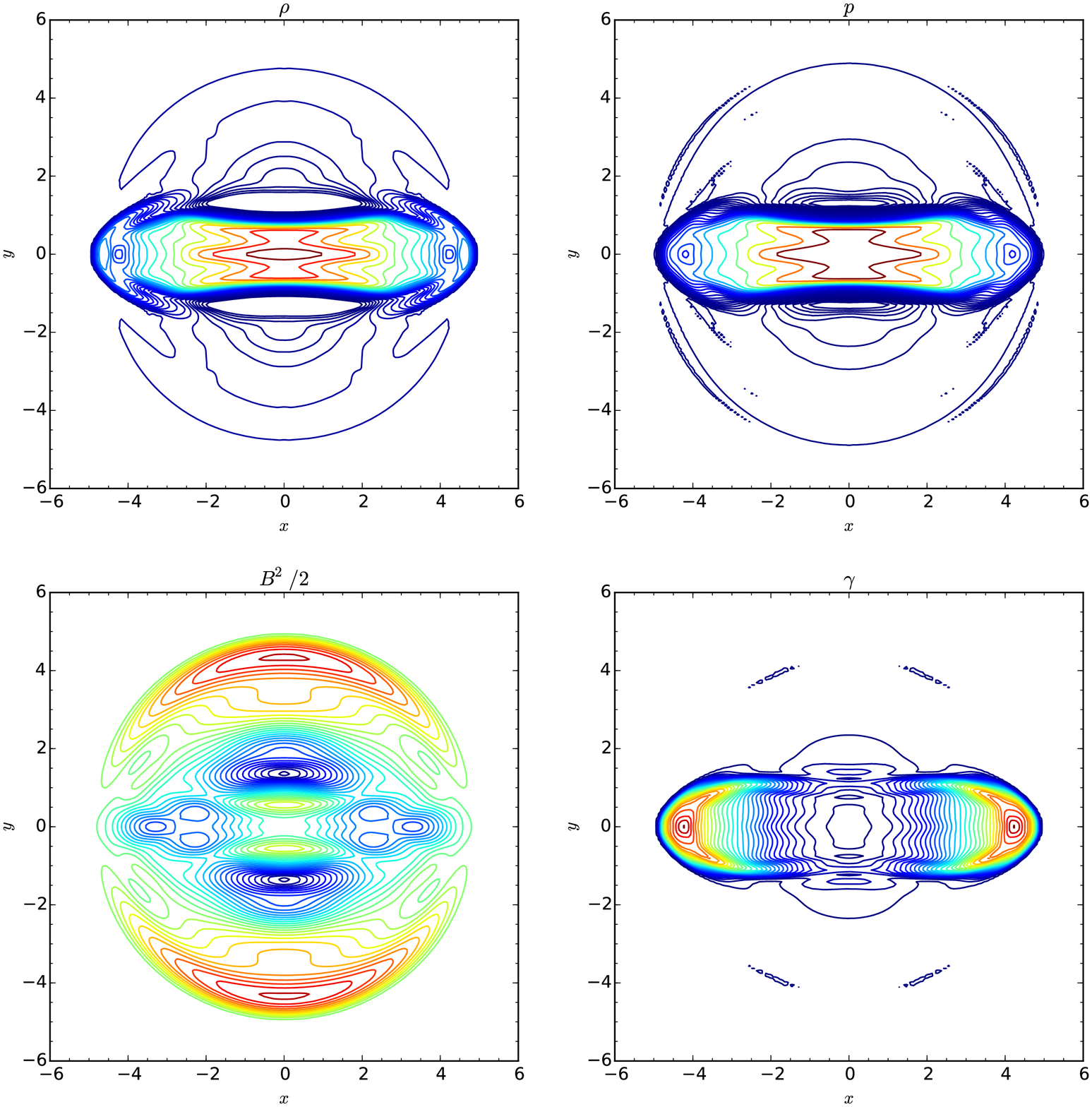}
 \else
 \begin{center}
  \includegraphics[scale=0.40]{figure/fig07.eps}
 \end{center}
 \fi

 \caption{Results for strong cylindrical explosion at $t = 4.0$ in a strongly
 magnetized medium ($B_0 = 1.0$). The format is the same as Figure 5. In each
 panel, 30 contours are drawn in the ranges $5.0 \times 10^{-5}$ to $2.0
 \times 10^{-3}$ for the density, $1.0 \times 10^{-4}$ to $1.0 \times 10^{-1}$
 for the gas pressure, $3.0 \times 10^{-1}$ to $7.0 \times 10^{-1}$ for the
 magnetic pressure, and $1.01$ to $3.0$ for the Lorentz factor,
 respectively. The contours are evenly spaced in logarithmic scale.}

 \label{fig:blast_strong}
\end{figure*}
}

\newcommand{\FigureEight}{
\begin{figure}[tb]
 \figurenum{8}
 \ifemulateapj
 \plotone{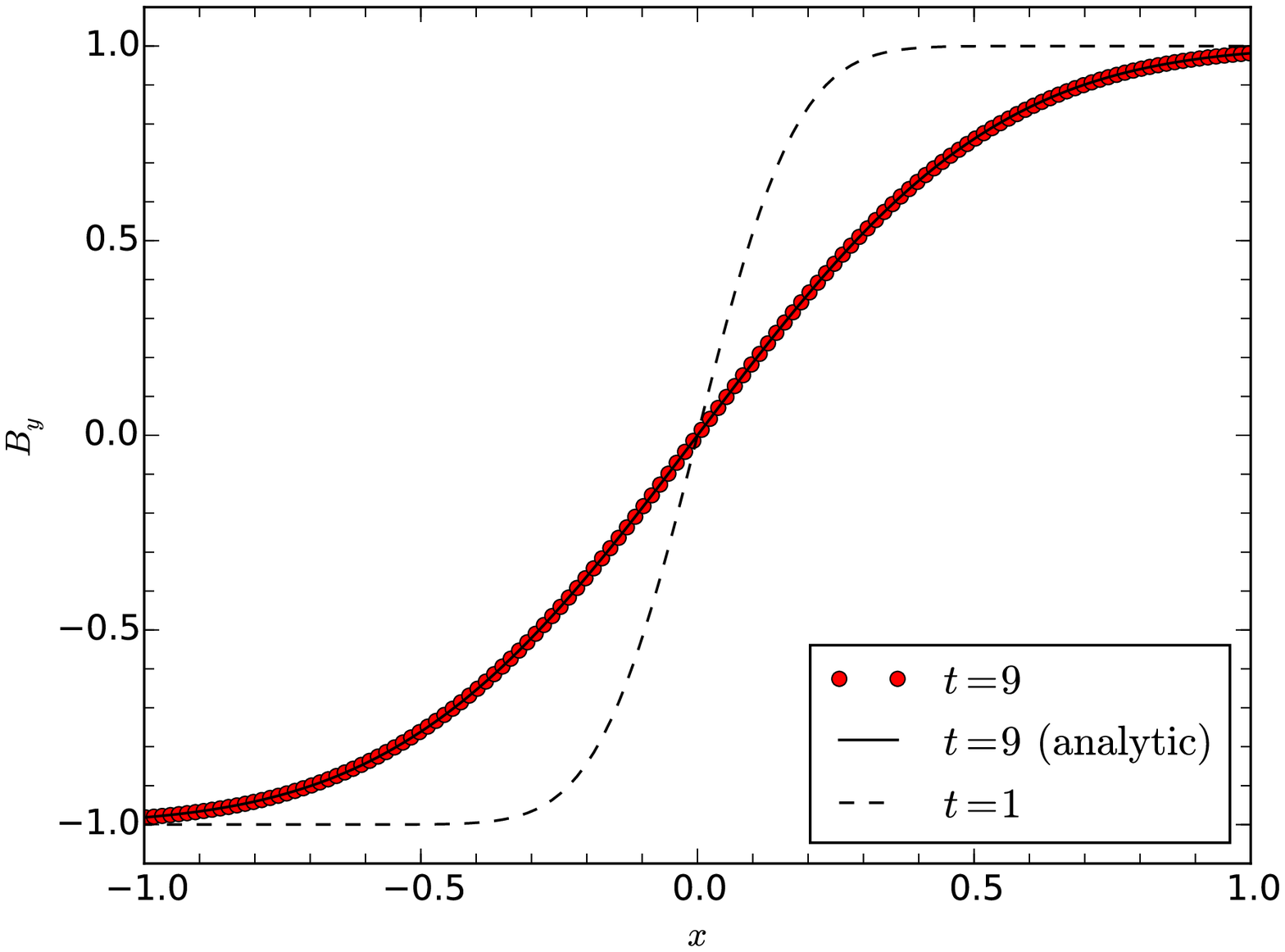}
 \else
 \begin{center}
  \includegraphics[scale=0.75]{figure/fig08.eps}
 \end{center}
 \fi

 \caption{Comparison between numerical and analytical solutions for
 self-similar current sheet problem. The profiles at $t = 9$ are shown with
 dots for the numerical solution, and in a solid line for the analytic solution,
 respectively. The initial condition ($t = 1$) is shown with dashed line.}

 \label{fig:selfsimilar}
\end{figure}
}

\newcommand{\FigureNine}{
\begin{figure}[tb]
 \figurenum{9}
 \ifemulateapj
 \plotone{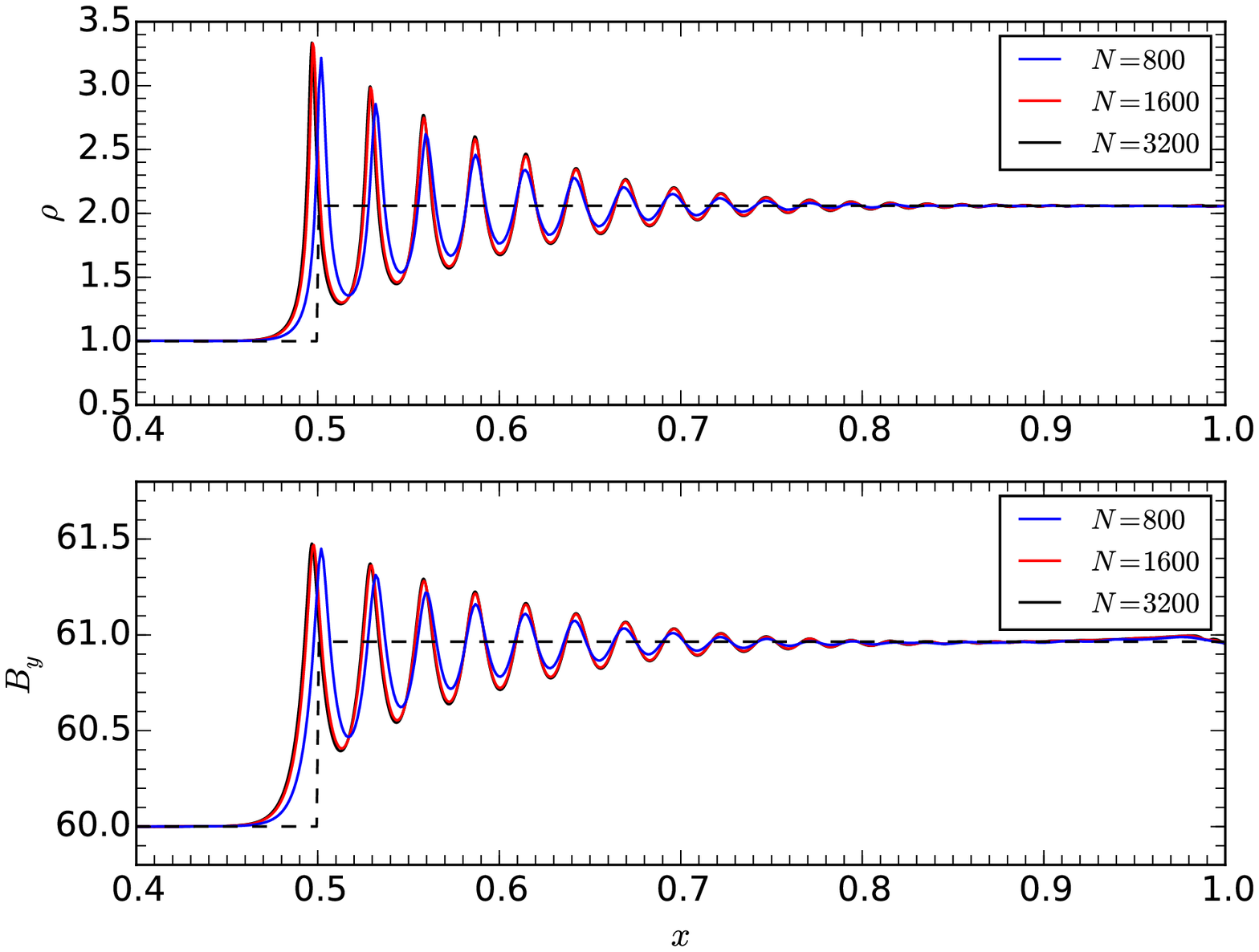}
 \else
 \begin{center}
  \includegraphics[scale=0.75]{figure/fig09.eps}
 \end{center}
 \fi

 \caption{Results for resistive perpendicular shock problem. The density (top)
 and $y$ component of the magnetic field (bottom) profiles are shown at $t =
 0.5$. The results with three different resolutions are shown with solid lines
 with different colors (blue, red, and black for $N = 800, 1600, 3200$,
 respectively). The dashed lines show the initial condition corresponding to
 the RMHD Rankine-Hugoniot relations.}

 \label{fig:perpshock}
\end{figure}
}

\newcommand{\FigureTen}{
\begin{figure}[tb]
 \figurenum{10}
 \ifemulateapj
 \plotone{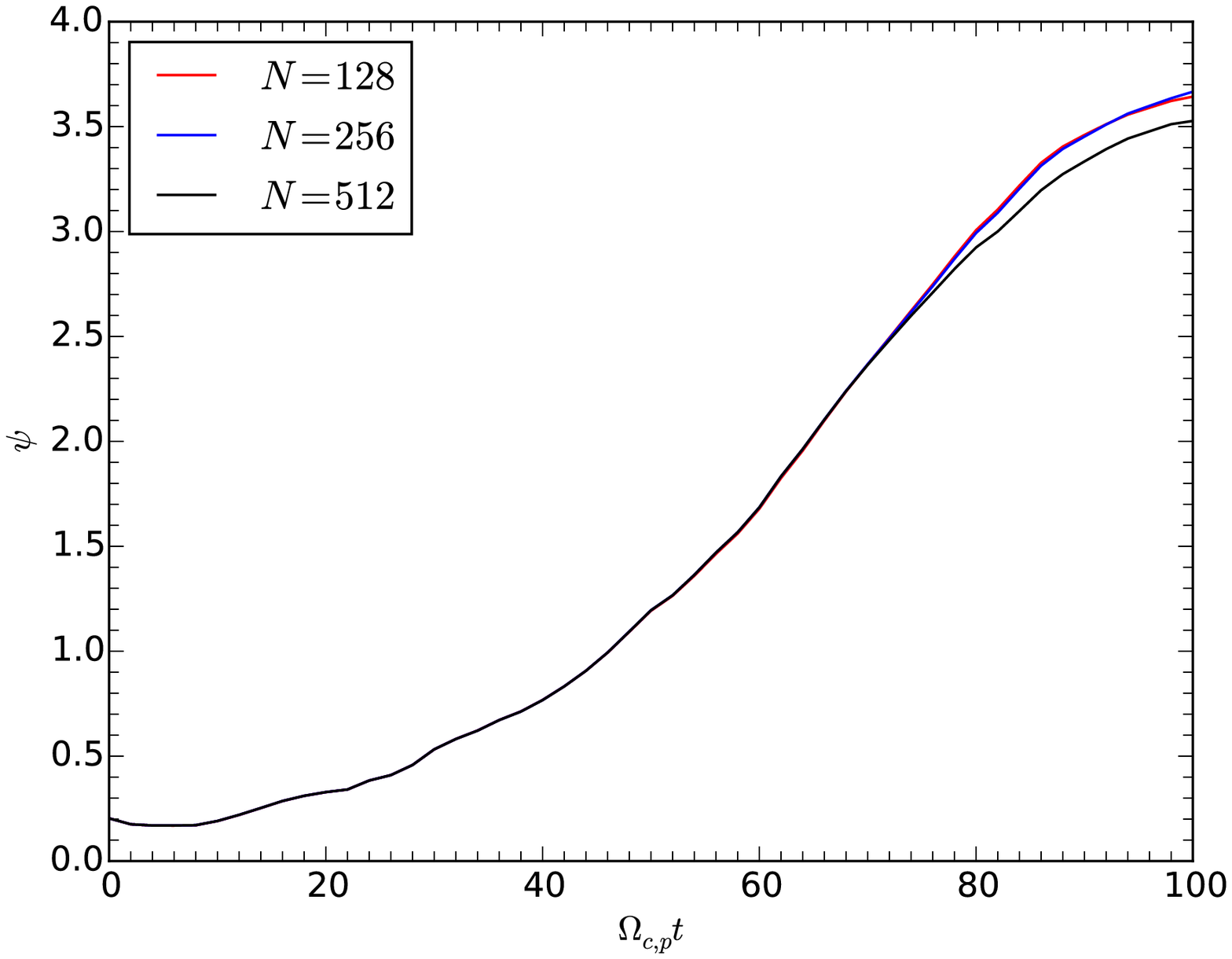}
 \else
 \begin{center}
  \includegraphics[scale=0.75]{figure/fig10.eps}
 \end{center}
 \fi

 \caption{Time development of reconnected magnetic flux for three runs with
 different resolutions $N = 128, 256, 512$.}

 \label{fig:recflux}
\end{figure}
}

\newcommand{\FigureEleven}{
\begin{figure*}[tb]
 \figurenum{11}
 \ifemulateapj
 \plotone{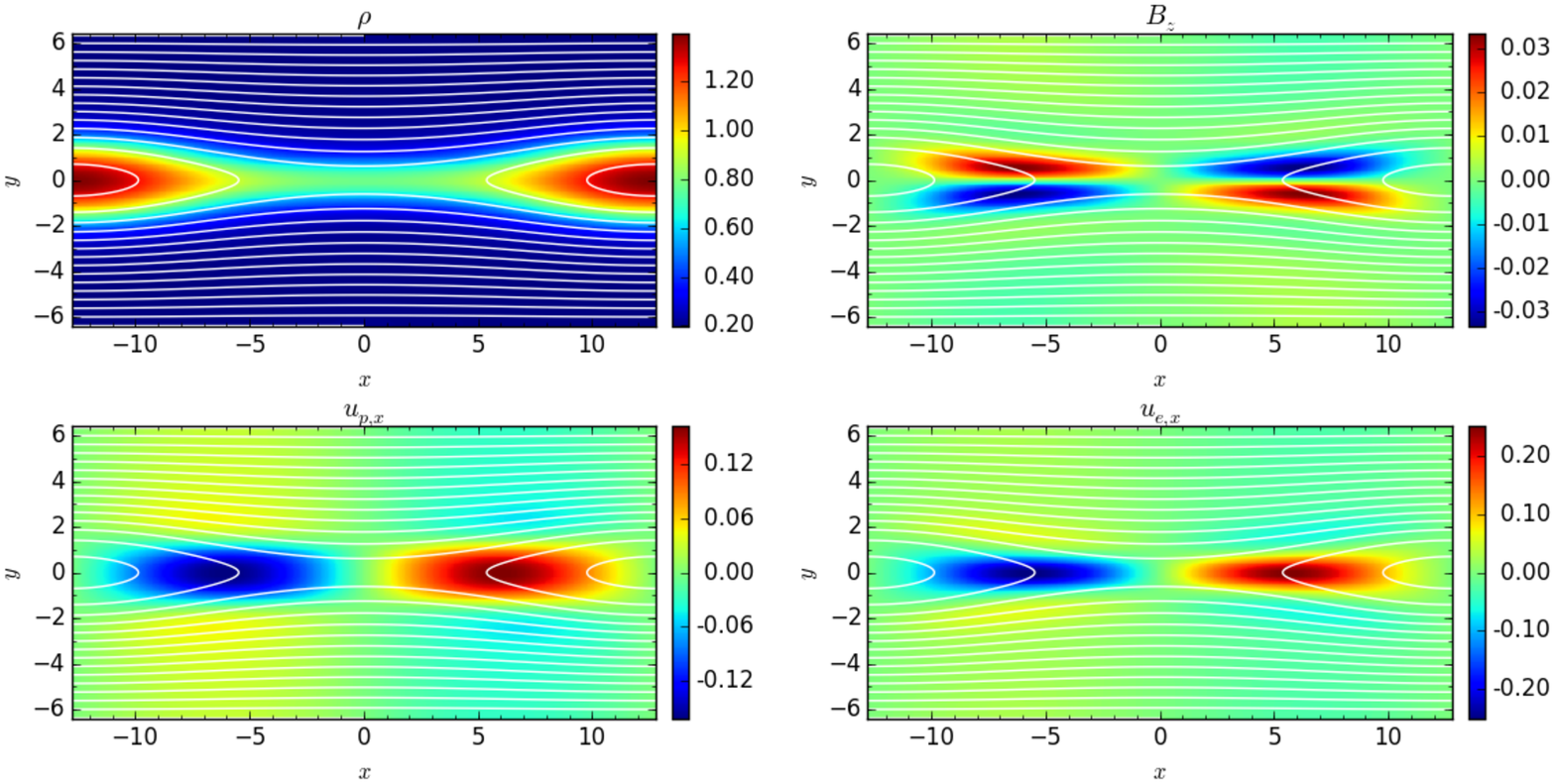}
 \else
 \begin{center}
  \includegraphics[scale=0.50]{figure/fig11.eps}
 \end{center}
 \fi

 \caption{Result for magnetic reconnection problem at $\Omega_{c,p} t =
 40$. The total mass density $\rho$ (top left), out-of-plane magnetic field
 $B_z$, $x$ component of four velocity for protons $u_{p,x}$ (bottom left),
 and electrons $u_{e,x}$ (bottom right) are shown. White contours represent
 the magnetic field lines.}

 \label{fig:mrx1}
\end{figure*}
}

\newcommand{\FigureTwelve}{
\begin{figure*}[tb]
 \figurenum{12}
 \ifemulateapj
 \plotone{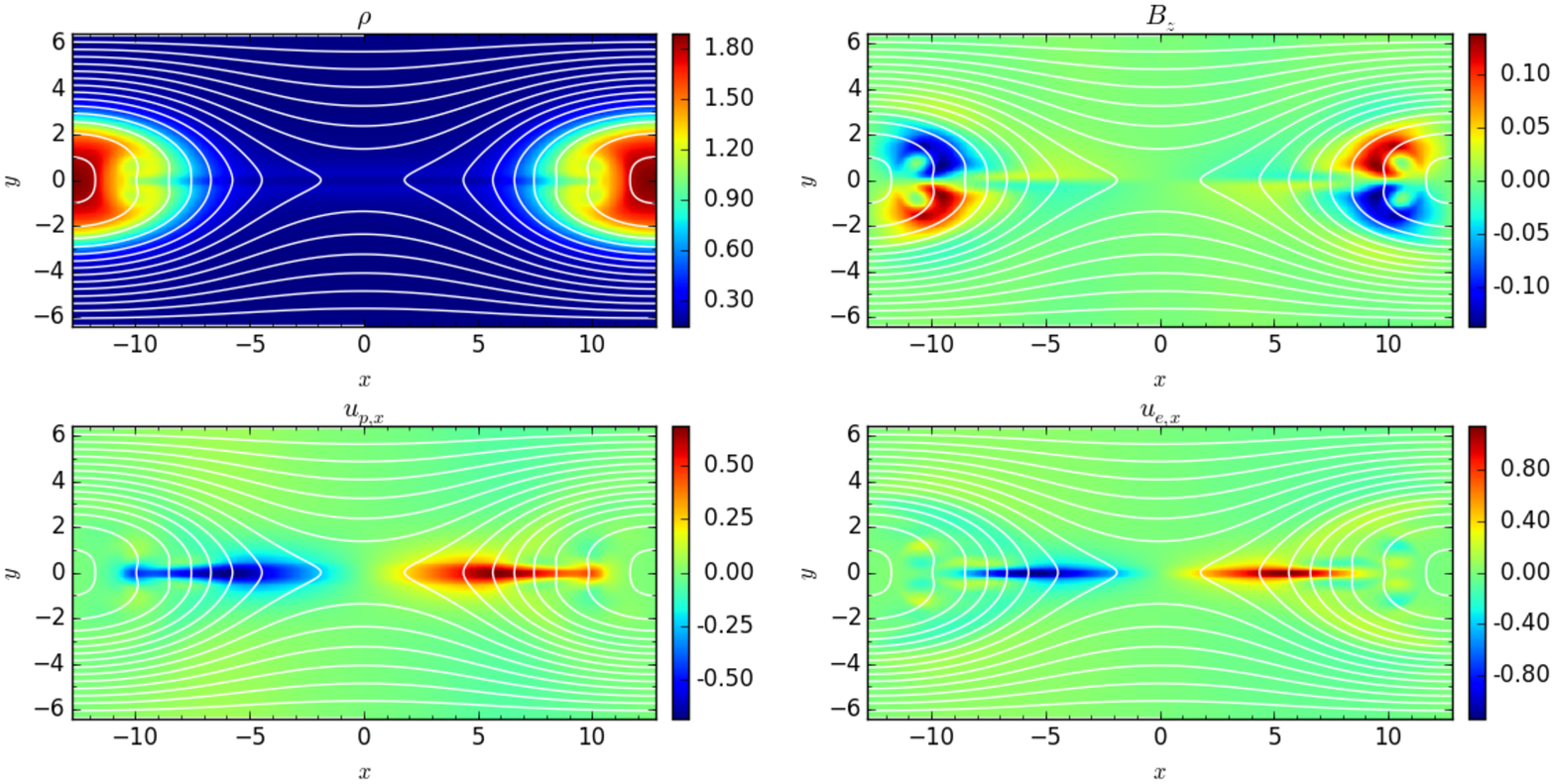}
 \else
 \begin{center}
  \includegraphics[scale=0.50]{figure/fig12.eps}
 \end{center}
 \fi

 \caption{Same as Fig.~\ref{fig:mrx1} but for $\Omega_{c,p} t = 80$.}

 \label{fig:mrx2}
\end{figure*}
}

\usepackage{amsmath,amssymb}

\shorttitle{Relativistic Two-fluid Electrodynamics Scheme}
\shortauthors{Amano}

\begin{document}

\title{A Second-order Divergence-constrained Multidimensional Numerical Scheme for
Relativistic Two-Fluid Electrodynamics}
\author{Takanobu Amano}
\email{amano@eps.s.u-tokyo.ac.jp} \affil{Department of Earth \& Planetary
Science, University of Tokyo, 113-0033, Japan}

\begin{abstract}
A new multidimensional simulation code for relativistic two-fluid
electrodynamics (RTFED) is described. The basic equations consist of the full
set of Maxwell's equations coupled with relativistic hydrodynamic equations
for separate two charged fluids, representing the dynamics of either an
electron-positron or an electron-proton plasma. It can be recognized as an
extension of conventional relativistic magnetohydrodynamics (RMHD). Finite
resistivity may be introduced as a friction between the two species, which
reduces to resistive RMHD in the long wavelength limit without suffering from
a singularity at infinite conductivity. A numerical scheme based on HLL
(Harten-Lax-Van Leer) Riemann solver is proposed that exactly preserves the
two divergence constraints for Maxwell's equations simultaneously. Several
benchmark problems demonstrate that it is capable of describing RMHD
shocks/discontinuities at long wavelength limit, as well as dispersive
characteristics due to the two-fluid effect appearing at small scales. This
shows that the RTFED model is a promising tool for high energy astrophysics
application.
\end{abstract}


\section{Introduction}
\label{intro}

There is growing evidence that the magnetic field plays a crucial role in
many astrophysical phenomena. In particular, it is now widely accepted that
relativistic outflows from compact objects involve magnetic fields. Examples
include pulsar winds, and jets from active galactic nuclei and gamma-ray
bursts. The magnetic field strength contained in the relativistic outflows may
be so strong that the dominant fraction of its energy flux is carried away in
the form of Poynting flux. Relativistic magnetohydrodynamics (RMHD) provides
the basic framework for understanding the dynamics of such highly magnetized
relativistic plasmas.

More importantly, such a Poynting-dominated outflow seems to be converted into
a matter-dominated state during its outward propagation. This indicates that
there must be an extremely efficient dissipation mechanism, converting the
magnetic field energy to particle kinetic energy. The best known example is
probably the $\sigma$-problem of the pulsar wind in the Crab nebula
\citep[e.g.,][]{1974MNRAS.167....1R,1984ApJ...283..694K}. Because the
relativistic wind itself is driven by the rapid rotation of a strongly
magnetized neutron star, it is likely to be in a Poynting-dominated regime at
least initially launched from the star. Observation of synchrotron emission
from the nebular region downstream of the termination shock, however,
indicates clearly that the energy density is dominated by relativistic
electron-positron pairs. Similar situations may occur in relativistic jets
from black holes
\citep[e.g.,][]{1994MNRAS.270..480T,2002A&A...391.1141D}. Although it is
plausible that the magnetic field plays the dominant role in accelerating the
relativistic jet \citep[e.g.,][]{1999ApJ...522..727K,2004ApJ...606..395M,
2005ApJ...625...60N,2006MNRAS.368.1561M,2007MNRAS.380...51K,2009MNRAS.394.1182K,
2009ApJ...699.1789T}, particle energization must be associated with
electromagnetic energy dissipation to explain the observed high energy
emission from relativistic particles.

It is thus natural to invoke magnetic reconnection as a possible mechanism to
explain the inferred dissipation in high energy astrophysical objects
\citep{1990ApJ...349..538C,1994ApJ...431..397M,2001ApJ...547..437L,
2003ApJ...591..366K}. Introducing finite resistivity is necessary to
take into account the effect of magnetic reconnection, which motivates the
development of resistive RMHD simulation codes and application to relativistic
magnetic reconnection
\citep[e.g.,][]{2006ApJ...647L.123W,2007MNRAS.374..415K,2010ApJ...716L.214Z,
2011ApJ...739L..53T,2013ApJ...775...50T}. On the other hand, Ohm's law in the
relativistic regime is not understood very well. Most of current resistive
RMHD simulations have adopted the simplest kind of Ohm's law without temporal
evolution terms. From numerical point of view, this form of Ohm's law has
difficulty at the infinite conductivity limit. Namely, there is an inherent
singularity at infinite conductivity (or zero resistivity), because the
numerical procedure involves division by resistivity. Intuitively, this sounds
odd, and a different form of resistivity may be preferred both physically and
numerically.

Another possible issue in an ideal RMHD is the assumption of a dense plasma,
such that any electric fields in the plasma rest frame are immediately
screened out. In a strongly magnetized regime, however, the number of charges
may not be enough to cancel the electric field. In other words, when the
magnetic field gradient requires, via Ampere's law, the current beyond the
limit that can be provided by the plasma, it must be compensated by the
displacement current, and there appears a finite (oscillatory) electric field
in the rest frame of the fluid. In terms of linear dispersion relation, these
waves correspond to high-frequency eigenmodes of a two-fluid plasma. The
electron plasma and/or cyclotron frequency is a typical threshold frequency,
above which the displacement current plays the dominant role. Such
high-frequency waves with phase speeds greater than the speed of light are
called superluminal waves, whereas all the RMHD waves have subluminal phase
speeds. Equivalently, RMHD is valid only when $B > E$ is satisfied, i.e., the
magnetic field is greater than the local electric field. In high-energy
astrophysics application, this condition may not necessarily be
satisfied. There has been a renewed interest in the possible importance of the
electric-field-dominated regime in the dynamics of a strongly magnetized
plasma \citep{2013ApJ...770...18A,2014MNRAS.442.2855T}, which is unfortunately
completely beyond the limit of RMHD.

These issues have motivated search for a better numerical framework other than
RMHD. Fully kinetic Particle-In-Cell (PIC) simulation is obviously the most
fundamental model, but it will not be an alternative to RMHD for realistic
modeling of macroscopic astrophysical phenomena, at least in the foreseeable
future. We believe that a relativistic two-fluid model coupled with the full
set of Maxwell's equations \citep[e.g.,][]{2009ApJ...696.1385Z} may possibly
be employed for this purpose. In this model, the relativistic hydrodynamic
(RHD) equations with the Lorentz force as an external force are solved
separately for electron and proton (or positron) fluids, respectively. The
dynamics is coupled through the electromagnetic field, which evolves according
to Maxwell's equations with the conduction current given by the moments of the
two fluids. It is essentially a fluid counterpart of PIC simulation, which
thus greatly reduces the computational requirement. As a compromise, it
ignores the kinetic effect such as collisionless wave-particle resonances and
finite Larmor radius effect, whereas dispersive characteristics arising from
the finite inertia of each species are correctly retained. Because the model
does not assume any relationships between the electric and magnetic fields, it
is possible to investigate the electric-field-dominated regime as
well. Although it is sometimes referred as relativistic two-fluid MHD, we
prefer to call the model as the relativistic two-fluid electrodynamics (RTFED)
to correctly represent its capability of describing the non-MHD regime $E >
B$. It can thus bridge the gap between the RMHD and electric-field-dominated
regimes.

Although application of the RTFED model to astrophysical problems so far has
been very limited at present \citep{2009ApJ...696.1385Z,
2009ApJ...705..907Z,2013ApJ...770...18A,2014MNRAS.438..704B,2016MNRAS.458.1939B},
we believe that it has the potential for more widespread use in the astrophysical
community. Part of the reason is that numerical methods that can be used for
the present system of equations have not adequately been explored. Although
many aspects of the equations are in common with, RHD, RMHD, as well as the
non-relativistic counterpart of two-fluid model
\citep[e.g.,][]{2011PhPl...18i2113S}, it is important to investigate
the applicability of existing numerical technologies to this particular set of
equations. We here present the application of the single state HLL (Harten-Lax-Van
Leer) approximate Riemann solver to the RTFED equations. Because
the electric field is also an independent variable that evolves in time
according to Ampere's law in this model, one has to be careful about the divergence
constraints of both the electric and magnetic fields in multidimensional
simulations. We have implemented a variant of the HLL-UCT (Upwind Constrained
Transport) scheme \citep{2004JCoPh.195...17L}, which was originally proposed
for MHD. It ingeniously combines the CT scheme \citep{1988ApJ...332..659E} and
a two-dimensional (2D) version of the HLL Riemann solver. We demonstrate that,
with a careful choice of collocation of physical quantities on a computational
mesh, the HLL-UCT works quite well for the RTFED equations. We believe
that this will be a better approach than the popular generalized Lagrangian
multiplier method \citep{2009ApJ...705..907Z,2014MNRAS.438..704B}, in
particular for the present set of equations. This is because a finite error in
Gauss's law for the electric field may couple with high-frequency Langmuir wave
fluctuations, and become a potential source of numerical instability. Our
numerical scheme is free this problem.

We also present a specific form of friction term between the species that, in
the long wavelength limit, reduces to the resistive Ohm's law that has
commonly been used in current resistive RMHD simulations. Our newly developed
three-dimensional (3D) simulation code does not assume any symmetry between
the two fluids, and can be used both for electron-positron or electron-proton
plasmas with an arbitrary mass ratio. Simulation results for several benchmark
problems are presented to demonstrate the robustness of our numerical
algorithm.

This paper is organized as follows. The basic equations are given in
section 2 with brief summary of the model characteristics. The numerical algorithm
is described in section 3, and numerical results for the test problems are shown
in section 4. Finally, section 5 gives conclusions of this paper.

\section{Simulation Model}
\label{model}
\subsection{Basic equations}
Throughout in this paper, we use the Lorentz-Heaviside units so that the
factor $\sqrt{4 \pi}$ does not appear in the basic equations. In other words,
the factor $1/\sqrt{4 \pi}$ is absorbed in the definition of the electric and
magnetic fields, whereas another factor $\sqrt{4 \pi}$ is absorbed in the
definition of the charge. The speed of light $c$ will appear explicitly for
the sake of completeness.

The RHD equations for a charged fluid under the influence of the electric and
magnetic fields ($\mathbf{E}$, $\mathbf{B}$) are given as follows:
\begin{align}
 &\frac{\partial}{\partial t}
 \left( \rho_s \gamma_s \right) +
 \nabla \cdot \left( \rho_s \mathbf{u}_s \right) = 0,
 \label{eq:mass} \\
 &\frac{\partial}{\partial t}
 \left( \frac{w_s}{c^2} \gamma_s \mathbf{u_s} \right) +
 \nabla \cdot
 \left( \frac{w_s}{c^2} \mathbf{u}_s \mathbf{u}_s + p_s \mathbf{I} \right) =
 \nonumber \\
 & \quad \mu_s \gamma_s \rho_s
 \left( \mathbf{E} + \frac{\mathbf{u}_s}{\gamma_s c} \times \mathbf{B} \right) +
 \mathbf{R}_s,
 \label{eq:momentum} \\
 &\frac{\partial}{\partial t}
 \left( w_s \gamma_s^2 - p_s \right) +
 \nabla \cdot \left( w_s \gamma_s \mathbf{u}_s \right) =
 \mu_s \rho_s \mathbf{u}_s \cdot \mathbf{E} +
 R^{0}_s.
 \label{eq:energy}
\end{align}
Here, the subscript $s$ denotes the particle species ($s = e$ for electrons
and $s = p$ for positrons or protons). The charge-to-mass ratio is denoted by
$\mu_s = e_s/m_s$, where $e_s$ and $m_s$ represent the charge and mass,
respectively. Because we assume that the plasma consists of either
electron-positron or electron-proton, the charge $e_s$ is either $+e$ or $-e$
where $e$ is the elementary charge. The fluid quantities are $\rho_s = n_s
m_s$, the proper mass density (with $n_s$ being the proper number density),
$(\gamma_s c, \mathbf{u}_s)$, the four velocity, and $p_s$, the proper gas
pressure. Throughout in this paper, we assume a polytropic equation of state
(EoS) with the ratio of specific heats denoted by $\Gamma$ (independent of the
particle species). Hence, the enthalpy density is given by $w_s = \rho_s c^2 +
\Gamma/(\Gamma-1) p_s$. We have also introduced a four vector $R^{\nu}_s =
(R^{0}_s/c, \mathbf{R}_s)$ which represents a frictional force operating
between the two fluids. We only assume anti-symmetry relationship $R^{\nu}
\equiv R^{\nu}_{p} = -R^{\nu}_e$ for the conservation of the total energy and
momentum, and the specific form of the friction term will be given later.

For later convenience, we introduce the definition for the specific enthalpy
\begin{align}
 h_s = \frac{w_s}{\rho_s c^2} =
 1 + \frac{\Gamma}{\Gamma-1} \frac{T_s}{m_s c^2},
 \label{eq:enthalpy}
\end{align}
which represents the increase of the fluid inertia relative to the rest-mass
density for a relativistically hot fluid $T_s/m_s c^2 \gtrsim 1$.

The two-fluid equations are coupled with each other through the
electromagnetic field that evolves according to Maxwell's equations:
\begin{align}
 & \frac{1}{c} \frac{\partial}{\partial t} \mathbf{E} =
 \nabla \times \mathbf{B} - \frac{\mathbf{J}}{c},
 \label{eq:dedt} \\
 & \frac{1}{c} \frac{\partial}{\partial t} \mathbf{B} =
 - \nabla \times \mathbf{E},
 \label{eq:dbdt} \\
 & \nabla \cdot \mathbf{E} = \varrho,
 \label{eq:dive} \\
 & \nabla \cdot \mathbf{B} = 0.
 \label{eq:divb}
\end{align}
The charge density $\varrho$ and the current density $\mathbf{J}$ are
respectively defined as follows
\begin{align}
 & \varrho = \sum \mu_s \gamma_s \rho_s,
 \\
 & \mathbf{J} = \sum \mu_s \rho_s \mathbf{u}_s.
\end{align}
The full set of Maxwell's equations involve the two divergence constraint
equations: Eqs.(\ref{eq:dive}) and (\ref{eq:divb}). Furthermore, the charge and
current densities must satisfy the charge continuity equation
\begin{align}
 \frac{\partial}{\partial t} \varrho + \nabla \cdot \mathbf{J} = 0
\end{align}
for consistency.

These equations provide a closed set of equations, which defines the governing
equations for RTFED. However, it is often convenient to rewrite the equations
into a form suitable for numerical treatment. As suggested by
\cite{2013ApJ...770...18A} \citep[see also,][]{1996MNRAS.279.1168M} for a pair
plasma, the sum of the two-fluid equations provides five conservation laws
that are very similar to RMHD, whereas the difference may be identified as the
charge conservation law and the generalized Ohm's law. For the latter set of
equations, we instead take the weighted sum with respect to the factor $\mu_s$
for generalization to an electron-proton plasma, or in fact, any two charged
fluids. We then obtain the following set of equations:
\begin{align}
 & \frac{\partial}{\partial t}
 \left( \sum \rho_s \gamma_s \right) +
 \nabla \cdot \left( \sum \rho_s \mathbf{u}_s \right) = 0.
 \label{eq:mass_sum} \\
 & \frac{\partial}{\partial t}
 \left( \sum \frac{w_s}{c^2} \gamma_s \mathbf{u}_s \right) +
 \nabla \cdot
 \left( \sum
 \left[ \frac{w_s}{c^2}\mathbf{u}_s\mathbf{u}_s + p_s \mathbf{I} \right]
 \right) =
 \nonumber \\
 &\quad \varrho \mathbf{E} + \frac{\mathbf{J}}{c} \times \mathbf{B}.
 \label{eq:momentum_sum} \\
 & \frac{\partial}{\partial t}
 \left( \sum \left[w_s \gamma_s^2 - p_s\right] \right) +
 \nabla \cdot \left( \sum w_s \gamma_s \mathbf{u}_s \right) =
 \mathbf{J} \cdot \mathbf{E},
 \label{eq:energy_sum} \\
 & \frac{\partial}{\partial t}
 \left( \sum \bar{\rho}_s \gamma_s \right) +
 \nabla \cdot \left( \sum \bar{\rho}_s \mathbf{u}_s \right) = 0.
 \label{eq:mass_dif} \\
 & \frac{\partial}{\partial t}
 \left( \sum \frac{\bar{w}_s}{c^2} \gamma_s \mathbf{u}_s \right) +
 \nabla \cdot
 \left( \sum
 \left[
 \frac{\bar{w}_s}{c^2}\mathbf{u}_s\mathbf{u}_s + \bar{p}_s \mathbf{I}
 \right]
 \right) =
 \nonumber \\
 &\quad \omega_{p}^2 \left(
 \gamma \mathbf{E} + \frac{\mathbf{u}}{c} \times \mathbf{B}
 \right) + \sum \mu_s \mathbf{R}_s
 \label{eq:momentum_dif} \\
 & \frac{\partial}{\partial t}
 \left( \sum \left[\bar{w}_s \gamma_s^2 - \bar{p}_s\right] \right) +
 \nabla \cdot \left( \sum \bar{w}_s \gamma_s \mathbf{u}_s \right) =
 \nonumber \\
 &\quad \omega_p^2 \mathbf{u} \cdot \mathbf{E} +
 \sum \mu_s R^{0}_s.
 \label{eq:energy_dif}
\end{align}
Henceforth, the sum must be taken over the two-fluid species unless explicitly
specified. Here we have introduced the notation: $\bar{\rho_s} = \mu_s
\rho_s$, $\bar{w}_s = \mu_s w_s$, $\bar{p}_s = \mu_s p_s$, $\omega_p^2 = \sum
\omega_{p,s}^2 = \sum \mu_s^2 \rho_s$. Notice that $\omega_{p,s}$ gives the
plasma frequency for each species as defined with the proper density, and
$\omega_p$ corresponds to the total plasma frequency. We define the average
four velocity $u^{\mu} = (\gamma c, \mathbf{u})$ (i.e., without subscript for
species) as follows
\begin{align}
 \gamma = \frac{\sum \mu_s^2 \rho_s \gamma_s}{\omega_p^2},
 \label{eq:average_g}
 \\
 \mathbf{u} = \frac{\sum \mu_s^2 \rho_s \mathbf{u}_s}{\omega_p^2},
 \label{eq:average_u}
\end{align}
which may be understood as the weighted average with respect to the proper
plasma frequency for each species. The
Eqs.~(\ref{eq:mass_sum})-(\ref{eq:energy_sum}) resulting from the (unweighted)
sum give the conservation laws for the total mass, momentum, and energy,
respectively. On the other hand, it is easy to understand that
Eq.~(\ref{eq:mass_dif}) gives the charge conservation law. The remaining
Eqs.~(\ref{eq:momentum_dif}) and (\ref{eq:energy_dif}) give the generalized
Ohm's law and will be discussed later.

The Lorentz force in the right-hand sides of the momentum and energy
conservation laws can be further rewritten using Maxwell's equations to obtain
the fully conservative form:
\begin{align}
 & \frac{\partial}{\partial t}
 \left(
 \sum \frac{w_s}{c^2} \gamma_s \mathbf{u}_s + \frac{\mathbf{S}}{c^2}
 \right) +
 \nonumber \\
 &\quad \nabla \cdot
 \left( \sum
 \left[ \frac{w_s}{c^2}\mathbf{u}_s\mathbf{u}_s + p_s \mathbf{I} \right]
 - \mathbf{T}
 \right) = 0,
 \\
 & \frac{\partial}{\partial t}
 \left( \sum \left[w_s \gamma_s^2 - p_s\right] + \mathcal{E} \right) +
 \nonumber \\
 &\quad \nabla \cdot
 \left( \sum w_s \gamma_s \mathbf{u}_s + \mathbf{S} \right) = 0.
\end{align}
The electromagnetic energy $\mathcal{E}$, the Maxwell stress tensor
$\mathbf{T}$, and the Poynting flux $\mathbf{S}$ are defined as follows
\begin{align}
 & \mathcal{E} = \frac{1}{2} \left( \mathbf{E}^2 + \mathbf{B}^2 \right),
 \\
 & \mathbf{T} =
 \mathbf{E} \mathbf{E} + \mathbf{B} \mathbf{B}
 - \mathcal{E} \mathbf{I},
 \\
 & \mathbf{S} = c \mathbf{E} \times \mathbf{B}.
\end{align}
Adopting this form of equations in numerical calculations with a conservative
scheme automatically guarantees the conservation of total energy and
momentum. We take this approach in this paper. Alternatively, one may also use
a semi-conservative form with source terms of Eqs.~(\ref{eq:momentum_sum}) and
(\ref{eq:energy_sum}). This approach may be useful to simulate
extremely low $\beta$ plasmas (where $\beta$ is the thermal to magnetic
pressure ratio), such as the pulsar magnetosphere where the force-free
approximation may be reasonable. In this case, the energy density of the
electromagnetic field is so large compared to the matter energy density,
that numerical stability issue may arise in the fully conservative
formulation.

Now we discuss for the generalized Ohm's law given by
Eqs.~(\ref{eq:momentum_dif}) and (\ref{eq:energy_dif}). The left-hand sides of
these equations describe the two-fluid effect arising from the difference
between the two species. They thus disappear in the low frequency limit, in
which case we obtain
\begin{align}
 &\omega_p^2 \left(
 \gamma \mathbf{E} + \frac{\mathbf{u}}{c} \times \mathbf{B}
 \right) +
 \sum \mu_s \mathbf{R}_s = \mathbf{0},
 \nonumber \\
 \quad
 &\omega_p^2
 \mathbf{u} \cdot \mathbf{E} +
 \sum \mu_s R^{0}_s = 0.
 \label{eq:mhd_limit}
\end{align}
In the ideal MHD limit, Ohm's law may be written as $\mathbf{E} + \mathbf{V}/c
\times \mathbf{B} = 0$ (where $\mathbf{V}$ is the three velocity of an MHD
flow). We thus find that the corresponding velocity is given by $\mathbf{V} =
\mathbf{u}/\gamma$. Recall that it is not the center-of-mass velocity, but
rather, a weighted average velocity with respect to the proper plasma
frequency $\omega_{p,s}$.

Although Ohm's law for the relativistic regime with a finite resistivity still
remains a controversial issue, the following form of Ohm's law with a finite
scalar resistivity $\eta$ is commonly adopted in current resistive RMHD
studies
\begin{align}
 \gamma \mathbf{E} + \frac{\mathbf{u}}{c} \times \mathbf{B} &=
 \eta \left( \mathbf{J} - \varrho_0 \mathbf{u} \right),
 \label{eq:ohm_mhd1}\\
 \mathbf{u} \cdot \mathbf{E} &=
 \eta \left( \varrho - \varrho_0 \gamma \right) c^2,
 \label{eq:ohm_mhd2}
\end{align}
where $\varrho_0 = \gamma \varrho - \mathbf{J} \cdot \mathbf{u} / c^2$ gives
the charge density as measured in the frame where $\mathbf{u} = 0$. We intend
to define the frictional four vector $R^{\mu}$ such that the generalized Ohm's
law reduces to Eqs.~(\ref{eq:ohm_mhd1}) and (\ref{eq:ohm_mhd2}) in the vanishing
two-fluid effect limit. This yields
\begin{align}
 \mathbf{R} &= - \eta \frac{\omega_p^2}{\mu_p - \mu_e}
 \left( \mathbf{J} - \varrho_0 \mathbf{u} \right),
 \\
 R^0 &= - \eta \frac{\omega_p^2}{\mu_p - \mu_e}
 \left( \varrho - \varrho_0 \gamma \right) c^2.
\end{align}
From this definition it is clear that $R^{\mu}$ is actually a four vector because
it is defined by the linear combination of the two four vectors: $u^{\mu}$,
$J^{\mu}$.

In this paper, we adopt this form of Ohm's law mainly because it makes
comparison with published MHD results easier. Note that this is valid even for
an electron-proton plasma regardless of the mass ratio. In the resistive RMHD
model, the current density determined from the Ohm's law
Eq.~(\ref{eq:ohm_mhd1}) is used for the time integration of Ampere's law. This
strategy has the fundamental difficulty in handling infinite conductivity
$\eta \rightarrow 0$ in the framework of resistive RMHD because calculation of
the current density involves division by $\eta$. In contrast, in the present
model, the current density is computed from the two-fluid moment quantities
directly rather than being determined from Ohm's law. Therefore, infinite
conductivity $\eta \rightarrow 0$ does not pose any numeral problems.

The friction term $R^{\mu}$ used in this paper differs from that used in the
earlier studies adopting the same set of equations
\citep[e.g.,][]{2009ApJ...705..907Z,2009ApJ...696.1385Z,2014MNRAS.438..704B}.
However, it is readily seen that our model reduces to the earlier work for a
perfectly symmetric case, i.e., a pair plasma with $\rho_p = \rho_e$, and
$\varrho_0 = 0$. This does not hold obviously in the presence of
charge-density fluctuations ($\rho_p \neq \rho_e$) even for a pair
plasma. Although it is natural to assume $\varrho_0 = 0$ on average for a pair
plasma, this will not be valid for a non-symmetric case. For consistency with
the RMHD Ohm's law, one has to define the fluid ``rest frame'' in which the
charge density $\varrho_0$ should be measured. However, the ``rest frame'' is
not a well-defined concept in the two-fluid regime. This is probably the
reason why application of the relativistic two-fluid model has been so far
limited to a pair plasma.

In contrast, the four velocity $u^{\mu}$ appearing in the Ohm's law defines a
natural reference frame where (under the low frequency limit) the magnetic
field line is at rest. This thus generalizes the convection velocity of the
magnetic field line to the two-fluid regime. By measuring $\varrho_0$ in this
particular frame, the consistency with the RMHD Ohm's law is assured. The
present form of Ohm's law can be understood as a straightforward extension of
non-relativistic version of the generalized Ohm's law
\citep{1978JCoPh..29..219H,2015JCP}, where the magnetic field convection
velocity is given in a similar form. As in the case of non-relativistic
situations, in a quasi-neutral electron-proton plasma, the single fluid RMHD
(or center-of-mass) flow velocity is dominated by the proton, whereas the
magnetic field convection is primarily governed by the electrons. This,
however, may not necessarily hold in a non-neutral region in a relativistic
plasma. Although we limit ourselves to discussion for a two-fluid plasma, it
should be easy to extend the analysis to a multifluid plasma model.

In summary, we use the following set of equations for the hydrodynamic part in
numerical simulations:
\begin{align}
 \frac{\partial}{\partial t} \mathbf{U} +
 \frac{\partial}{\partial x} \mathbf{F} +
 \frac{\partial}{\partial y} \mathbf{G} +
 \frac{\partial}{\partial z} \mathbf{H} =
 \mathbf{Q},
 \label{eq:fluid}
\end{align}
where $\mathbf{U}$ denotes the conservative variables defined as
\begin{align}
 \renewcommand*{\arraystretch}{1.5}
 \mathbf{U} (\mathbf{W}, \mathbf{E}, \mathbf{B}) =
 \begin{pmatrix}
  \displaystyle
  \sum \rho_s \gamma_s
  \\ \displaystyle
  \sum \frac{w_s}{c^2} \gamma_s u_{s,x} + \frac{S_x}{c^2}
  \\ \displaystyle
  \sum \frac{w_s}{c^2} \gamma_s u_{s,y} + \frac{S_y}{c^2}
  \\ \displaystyle
  \sum \frac{w_s}{c^2} \gamma_s u_{s,z} + \frac{S_z}{c^2}
  \\ \displaystyle
  \sum \left[w_s \gamma_s - p_s \right] + \mathcal{E}
  \\ \displaystyle
  \sum \bar{\rho}_s \gamma_s
  \\ \displaystyle
  \sum \frac{\bar{w}_s}{c^2} \gamma_s u_{s,x}
  \\ \displaystyle
  \sum \frac{\bar{w}_s}{c^2} \gamma_s u_{s,y}
  \\ \displaystyle
  \sum \frac{\bar{w}_s}{c^2} \gamma_s u_{s,z}
  \\ \displaystyle
  \sum \left[ \bar{w}_s \gamma_s - p_s \right]
 \end{pmatrix},
\end{align}
while $\mathbf{W} = \left\{ \rho_p, u_{p,x}, u_{p,y}, u_{p,z}, p_p, \rho_e,
u_{e,x}, u_{e,y}, u_{e,z}, p_e \right\}$ represents the two-fluid primitive
variables.

The flux in the $x$ direction $\mathbf{F}$ and the source term $\mathbf{Q}$ are
given by
\begin{align}
 \renewcommand*{\arraystretch}{1.5}
 &\mathbf{F} (\mathbf{W}, \mathbf{E}, \mathbf{B}) =
 \begin{pmatrix}
  \displaystyle
  \sum \rho_s u_{s,x}
  \\ \displaystyle
  \sum \left[
  \frac{w_s}{c^2} u_{s,x}^2 + p_s \right] - T_x
  \\ \displaystyle
  \sum \frac{w_s}{c^2} u_{s,x} u_{s,y} - T_y
  \\ \displaystyle
  \sum \frac{w_s}{c^2} u_{s,x} u_{s,z} - T_z
  \\ \displaystyle
  \sum w_s \gamma_s u_{s,x} + S_x
  \\ \displaystyle
  \sum \bar{\rho}_s u_{s,x}
  \\ \displaystyle
  \sum
  \left[ \frac{\bar{w}_s}{c^2} u_{s,x}^2 + \bar{p}_s \right]
  \\ \displaystyle
  \sum \frac{\bar{w}_s}{c^2} u_{s,x} u_{s,y}
  \\ \displaystyle
  \sum \frac{\bar{w}_s}{c^2} u_{s,x} u_{s,z}
  \\ \displaystyle
  \sum \bar{w}_s \gamma_s u_{s,x}
 \end{pmatrix},
 \\
 &\mathbf{Q} (\mathbf{W}, \mathbf{E}, \mathbf{B})=
 \nonumber \\
 &\quad \omega_p^2
 \begin{pmatrix}
  \displaystyle
  0
  \\ \displaystyle
  0
  \\ \displaystyle
  0
  \\ \displaystyle
  0
  \\ \displaystyle
  0
  \\ \displaystyle
  0
  \\ \displaystyle
  \gamma E_x + \left(u_y B_z - u_z B_y\right)/c
  - \eta \left(J_x - \varrho_0 u_x\right)
  \\ \displaystyle
  \gamma E_y + \left(u_z B_x - u_x B_z\right)/c
  - \eta \left(J_y - \varrho_0 u_y\right)
  \\ \displaystyle
  \gamma E_z + \left(u_x B_y - u_y B_x\right)/c
  - \eta \left(J_z - \varrho_0 u_z\right)
  \\ \displaystyle
  u_x E_x + u_y E_y + u_z E_z
  - \eta \left(\varrho - \varrho_0 \gamma\right) c
 \end{pmatrix}.
\end{align}
The fluxes in the $y$ and $z$ directions $\mathbf{G}$, and $\mathbf{H}$ are
similarly obtained by cyclic permutation of indices. Time evolution of the
electromagnetic field $\mathbf{E}$, $\mathbf{B}$ is computed using
the full set of Maxwell's equations Eqs.~(\ref{eq:dedt}) and (\ref{eq:dbdt})
under the constraints from Eqs.~(\ref{eq:dive}) and (\ref{eq:divb}).

\subsection{Model Characteristics}
The RTFED equations involve sixteen 16 with two divergence
constraints. This indicates that 14 wave modes exist in the system. Two
of them are entropy modes for each fluid, which are however tightly coupled to
give a degenerated (standard hydrodynamics) entropy mode in most
circumstances. For categorizing the rest of wave modes, it is convenient to
consider strictly parallel propagation with respect to the ambient magnetic
field. The longitudinal waves involving density perturbations are called
Langmuir and ion-acoustic waves, respectively. The former is characterized by
charge-density fluctuations, whereas the latter is essentially a quasi-neutral
mode. The transverse modes may be divided into subluminal ($\omega/k < c$) and
superluminal ($\omega/k > c$) waves in terms of their phase speeds with
respect to the speed of light. Alfv\'enic waves are those with subluminal
phase speeds, whereas electromagnetic waves are superluminal. Each of
transverse waves have two polarizations (i.e., right- or left-handed
polarization). Therefore, the transverse waves give four different wave
modes. Each of these six non-zero frequency modes can propagate both in
positive and negative directions. Taking into account the two entropy modes,
the number of modes adds up to 14 in total.

The typical time scale for each fluid is represented by the inverse proper
plasma frequency $1/\omega_{p,s}$. The fluid response changes significantly in
between fast and slow time scale phenomena with respect to $1/\omega_{p,s}$.
For frequencies much lower than than the plasma frequency of both species (or
$\omega \ll \omega_{p}$), the two fluids essentially move together and
one-fluid approximation becomes appropriate. Actually,
Eq.~(\ref{eq:mhd_limit}) was obtained in the limit $\omega / \omega_{p}
\rightarrow 0$, which indicates that the plasma is frozen-in to the magnetic
field line motion.

The corresponding spatial scale given by $c / \omega_{p,s}$ provides the skin
depth, representing the typical scale length at which the dispersive effect
appears in the RMHD normal mode. High-frequency (Langmuir and electromagnetic)
waves that do not exist in RMHD also change their character at the same
spatial scale. For sufficiently small wavenumber $k c / \omega_{p} \rightarrow
0$, the high-frequency waves are cut off around the plasma frequency (actual
cut-off frequency depends on polarization and magnetic field strength) but
continue to exist. Therefore, even in the long wavelength limit, the
eigenmodes consist of the standard RMHD modes and high-frequency plasma
waves. The presence of the high-frequency waves actually imposes a severe
restriction on the time step of explicit time integration schemes.

Another typical time scale for a magnetized plasma is given by the inverse
cyclotron frequency $1/\Omega_{c,s} = 1 / (\mu_s B / c)$ with the magnetic
field strength $B$ measured in the rest frame of the fluid. Here we introduce
the magnetization parameter as the ratio between the cyclotron and plasma
frequency squared: $\sigma_{s} \equiv \Omega_{c,s}^2/\omega_{p,s}^2 =
B^2/\rho_s c^2$, which can also be understood as the ratio between the
magnetic field and rest-mass energy densities. One may think that a charged
fluid with $\sigma_{s} \gtrsim 1$ is strongly magnetized. In this case, the
skin depth is longer than the Larmor radius, which gives the typical scale
length for the kinetic effect. Therefore, we can naively guess that the RTFED
model is better suited for this case because the kinetic effect is expected to
be less important. Conversely, the model loses its strict validity at scale
length comparable to the skin depth for a weakly magnetized plasma because of
the lack of the kinetic effect.

Similarly, the Debye length defined with the thermal velocity $u_{th,s}/c =
\sqrt{p_s / n_s m_s c^2}$ and the plasma frequency gives the length scale, at
which the Landau resonance against Langmuir and ion-acoustic waves becomes
important. Again, the model under the fluid approximation is no longer valid
at this scale.

In the presence of finite resistivity, one can also define a frictional
relaxation time scale normalized to $1/\omega_p$ by $\tau_f = 1/\omega_p
\eta$. Alternatively, in the long wavelength limit, the magnetic field
evolution can be described by the diffusion equation with a diffusion
coefficient of $\eta c^2$.

%Finally, we note that the linear dispersion relation of RTFED equations with
%finite temperature is essentially unchanged from that of the cold two-fluid
%plasma. Finite temperature effect may be absorbed in the definition of the
%rest mass, by replacing it with the effective mass $m_s h_s$.

\section{Numerical Algorithm}
\label{algorithm}

\subsection{Primitive recovery}
It is well known that, in contrast to non-relativistic counterparts, both RHD
and RMHD codes involve numerical solution of nonlinear equations to recover
the primitive variables $\mathbf{W}$ from conservative variables
$\mathbf{U}$. Because Eq.~(\ref{eq:fluid}) is mathematically derived from the
original two-fluid equations written separately, the primitive recovery
problem for the present model is identical to that of RHD: Because the
electromagnetic field is directly obtained from Maxwell's equations, its
contribution may be subtracted from the conservative variables. Then, a simple
arithmetic calculation can separate conservative variables for each fluid.

The problem now is how to obtain $\rho, \mathbf{u}, p$ from the conservative
variables for the RHD equations: $D = \gamma \rho, M = \gamma w
|\mathbf{u}|/c^2, K = (\gamma^2 w - p)/c^2$. Here, we have omitted the
subscript for fluid species. We adopt a method specialized to the polytropic
EoS proposed by \cite{2009ApJ...696.1385Z}, which gives $|\mathbf{u}|$ as a
root of the following single quartic equation:
\begin{align}
 f(X) = X^4 + a X^3 + b X^2 + c X + d = 0,
\end{align}
where
\begin{align}
 \Theta &= \Gamma/(\Gamma-1), \quad Y = M/K, \quad Z = D/K,
 \\
 a &=-2 Y Z/(\Theta (1+Y)(1-Y)),
 \\
 b &= (\Theta^2 - 2 \Theta (\Theta-1) Y^2 - Z^2)/(\Theta^2(1+Y)(1-Y)),
 \\
 c &=-(2(\Theta-1) Y Z)/(\Theta^2(1+Y)(1-Y)),
 \\
 d &=-(\Theta-1)^2 Y^2/(\Theta^2(1+Y)(1-Y)).
\end{align}
Other quantities are easily computed once the root for $f(|\mathbf{u}|) = 0$
is obtained. Following \cite{2009ApJ...696.1385Z}, we use the Brown method and
solve the quartic equation analytically. We prefer this method because it
ensures that the physical root is always obtained by a fixed number of
operation (i.e., without iteration) as long as the input is physically valid.

\subsection{Discretization in space}
The spatial discretization used in this paper is similar to typical CT-based MHD
codes. Namely, the magnetic field is collocated at the face center in the
normal direction (e.g., $B_x$ is defined at the center of the $x$-face)
whereas the fluid quantities are collocated at the cell center. This choice of
collocation is known to work quite well for MHD when a Riemann solver for the
cell-centered variables is combined with a suitable multidimensional flux for
the induction equation
\citep[e.g.,][]{2000ApJ...530..508L,2004JCoPh.195...17L,
2005JCoPh.205..509G,2007A&A...473...11D,2010JCoPh.229.1970B}. We have recently
demonstrated that essentially the same technique can be applied to a
non-relativistic quasi-neutral two-fluid model \citep{2015JCP}, indicating the
robustness of this approach.

\ifemulateapj
\FigureOne
\fi

In the RTFED equations, the electric field also evolves in time as a primary
variable according to Ampere's law. For consistency, it must satisfy Gauss's
law Eq.~(\ref{eq:dive}), yet another divergence constraint. Therefore, we have
to carefully design the collocation strategy. In the present study, we adopt
the collocation shown in Fig.~\ref{fig:collocation}. Similarly to the magnetic
field, we define both the electric field and current density at the face
center. As we show below, this choice is well suited for the two
divergence constraints being preserved simultaneously within machine
precision.

Notice that this definition differs from the conventional Yee mesh, which is
often employed in Particle-In-Cell (PIC) simulations. It defines the electric
and magnetic fields in a fully staggered manner. In other words, if the
electric field is collocated on the face, the magnetic field should be
collocated on the edge. At the second-order level, the magnetic field can be
updated directly from the primary electric field by simply recognizing it as
the flux. In the absence of the conduction current, the anti-symmetry between
the electric and magnetic field suggests that the electric field can also be
updated in the same way.

In the presence of finite charge and current densities, these variables must
be defined to be consistent with the collocation of the electric field. In typical
PIC simulations, an appropriate charge-conservative particle deposition scheme
\citep{2001CoPhC.135..144E} is employed to satisfy the charge continuity
equation. By defining the current density on the same collocation as the
electric field of the Yee mesh, the electromagnetic field can be evolved in a
manner that is fully consistent with both of the divergence constraints.

On the other hand, the Yee mesh is not a convenient choice when it applies to
the RTFED equations, because we intend to update the fluid part by using a
Riemann solver. This requires the collocation such that the fluid quantities
are defined at the cell center, whereas the magnetic field is defined at the
face center. Thus, it is natural to define the charge and current densities at
the cell center and face center, respectively. (Recall the charge continuity
equation comes from the mass conservation law of the two fluids.) Therefore,
in order to satisfy Gauss's law Eq.~(\ref{eq:dive}), it is also better to
define the primary electric field also at the face center.

With this consideration, we believe that our symmetric collocation of the
electromagnetic field would be the best choice to preserve the two divergence
constraints at the same time. The numerical experiments shown in section 4
actually support this conjecture.

\subsection{Interpolation}
The primary electromagnetic field is defined at the center of the normal face,
whereas the Riemann solver at the face also requires the transverse
components. For simplicity, we here use a 1D interpolation as described
below. At second-order accuracy, the simple arithmetic mean of the two
face-centered values is sufficient to obtain the cell-centered electric field:
\begin{align}
 &E_{x;i,j,k} = \frac{1}{2} \left( E_{x;i+1/2,j,k} + E_{x;i-1/2,j,k} \right) \\
 &E_{y;i,j,k} = \frac{1}{2} \left( E_{y;i,j+1/2,k} + E_{y;i,j-1/2,k} \right),\\
 &E_{z;i,j,k} = \frac{1}{2} \left( E_{z;i,j,k+1/2} + E_{y;i,j,k-1/2} \right),
\end{align}
and the cell-centered magnetic field may also be computed in the same way.

When higher-order accuracy is desired, the interpolation should also be
replaced with higher order ones accordingly. Not only that, it would be better
to use a multidimensional interpolation scheme that is consistent with the
divergence constraints. A multidimensional divergence-free reconstruction
scheme for the magnetic field in the entire control volume was presented by
\cite{2001JCoPh.174..614B,2004ApJS..151..149B,2009JCoPh.228.5040B}. The method
has recently been extended to the reconstruction of the electric field, in
which the presence of a finite charge-density profile is taken into account
\citep{2016JCoPh.318..169B}.

\subsection{Update for cell-centered variables}
The conservation law for the fluid quantities defined at the cell center may
be discretized into the following semidiscrete form
\begin{align}
 \frac{d}{d t} \mathbf{U}_{i,j,k} =
 &- \frac{1}{\Delta x}
 \left(
 \hat{\mathbf{F}}_{i+1/2,j,k} - \hat{\mathbf{F}}_{i-1/2,j,k}
 \right)
 \nonumber \\
 &- \frac{1}{\Delta y}
 \left(
 \hat{\mathbf{G}}_{i,j+1/2,k} - \hat{\mathbf{G}}_{i,j-1/2,k}
 \right)
 \nonumber \\
 &- \frac{1}{\Delta z}
 \left(
 \hat{\mathbf{H}}_{i,j,k+1/2} - \hat{\mathbf{H}}_{i,j,k-1/2}
 \right)
 + \mathbf{Q}_{i,j,k},
\end{align}
where $\Delta x$, $\Delta y$, $\Delta z$ are the grid sizes in each direction,
and $\hat{\mathbf{F}}_{i+1/2,j,k}$, $\hat{\mathbf{G}}_{i,j+1/2,k}$,
$\hat{\mathbf{H}}_{i,j,k+1/2}$ are the numerical fluxes. Below we describe the
algorithm to evaluate the flux in the $x$ direction based on the HLL Riemann
solver. Because we consider only a second-order scheme in this paper, the
one-dimensional (1D) flux evaluation procedure may be applied in a
dimension-by-dimension fashion to obtain the multidimensional fluxes. Note
that this approach is not valid for finite volume schemes of accuracy better
than second order, in which a fully multidimensional reconstruction including
cross terms is needed \citep[see,][]{2016JCoPh.318..169B}.

We introduce the following symbolic expression for the reconstruction
procedure:
\begin{align}
 f_{i+1/2,j,k}^{\rmR_x} \equiv {\mathcal R}^{\rmR_x} (f_{i,j,k}),
 \quad
 f_{i+1/2,j,k}^{\rmL_x} \equiv {\mathcal R}^{\rmL_x} (f_{i,j,k}).
\end{align}
This represents a 1D reconstruction in the $x$ direction to obtain, respectively,
the right ($f_{i+1/2,j,k}^{\rmR_x}$) and left ($f_{i+1/2,j,k}^{\rmL_x}$)
states at the $x$ face from the cell-centered variables. Similarly, 1D
reconstruction in the $y$ and $z$ directions are represented by the superscript
$\rmR_y$, $\rmL_y$, $\rmR_z$, $\rmL_z$.

We apply a piecewise linear reconstruction for the cell-centered fluid
primitive variables $\rho_p$, $\mathbf{u}_p$, $p_p$, $\rho_e$, $\mathbf{u}_e$,
$p_e$, as well as the {\it transverse} components of the electromagnetic field
$E_y, E_z, B_y, B_z$. The Monotonized Central (MC) slope limiter is used for
the reconstruction in this study. Recall that we do not need any
interpolation/reconstruction for the normal component of the electromagnetic
field $E_x$, $B_x$ because they are already defined at the face center as
primary variables. This removes the ambiguity in the definition of the normal
component, which is indeed consistent with the fact that the normal component
is constant in the 1D Riemann problem. To simplify the notation, we may write
$E^{\rmR_x}_{x;i+1/2,j,k} = E^{\rmL_x}_{x;i+1/2,j,k} = E_{x;i+1/2,j,k}$ and
$B^{\rmR_x}_{x;i+1/2,j,k} = B^{\rmL_x}_{x;i+1/2,j,k} = B_{x;i+1/2,j,k}$.

Using the above procedure gives both right ($\mathbf{W}^{\rmR_x}_{i+1/2,j,k}$,
$\mathbf{E}^{\rmR_x}_{i+1/2,j,k}$, $\mathbf{B}^{\rmR_x}_{i+1/2,j,k}$) and left
($\mathbf{W}^{\rmL_x}_{i+1/2,j,k}$, $\mathbf{E}^{\rmL_x}_{i+1/2,j,k}$,
$\mathbf{B}^{\rmL_x}_{i+1/2,j,k}$) states defined appropriately at each face.
Accordingly, we are ready to solve the Riemann problem. We employ the single
state HLL Riemann solver. This is certainly the simplest choice, but it is
also the only option at present. The eigenstructure of the RTFED equations
(eigenvalue problem for a $14\times14$ matrix) has not been analyzed so
far. Furthermore, the Riemann problem is no longer self-similar due to the
dispersive nature of the system. The HLL Riemann solver is thus a reasonable
compromise to avoid the complexity in the system. In reality, because the
maximum wave speed are always given by the speed of light $c$, the HLL Riemann
solver applied to this system automatically reduces to a global Lax-Friedrichs
scheme. Consequently, the numerical flux is given by
\begin{align}
 \mathbf{\hat{F}}_{i+1/2,j,k} =
 &\frac{1}{2}
 \left(
 \mathbf{F}^{\rmL}_{i+1/2,j,k} + \mathbf{F}^{\rmR}_{i+1/2,j,k}
 \right) -
 \nonumber \\
 &\frac{c}{2}
 \left(
 \mathbf{U}^{\rmR}_{i+1/2,j,k} - \mathbf{U}^{\rmL}_{i+1/2,j,k}
 \right),
 \label{eq:hll}
\end{align}
where
\begin{align}
 & \mathbf{F}^{\rmR} = \mathbf{F}(
 \mathbf{W}^{\rmR}_{i+1/2,j,k},
 \mathbf{E}^{\rmR}_{i+1/2,j,k},
 \mathbf{B}^{\rmR}_{i+1/2,j,k}),
 \, \nonumber \\
 & \mathbf{F}^{\rmL} = \mathbf{F}(
 \mathbf{W}^{\rmL}_{i+1/2,j,k},
 \mathbf{E}^{\rmL}_{i+1/2,j,k},
 \mathbf{B}^{\rmL}_{i+1/2,j,k}),
 \\
 & \mathbf{U}^{\rmR} = \mathbf{U}(
 \mathbf{W}^{\rmR}_{i+1/2,j,k},
 \mathbf{E}^{\rmR}_{i+1/2,j,k},
 \mathbf{B}^{\rmR}_{i+1/2,j,k}),
 \, \nonumber \\
 & \mathbf{U}^{\rmL} = \mathbf{U}(
 \mathbf{W}^{\rmL}_{i+1/2,j,k},
 \mathbf{E}^{\rmL}_{i+1/2,j,k},
 \mathbf{B}^{\rmL}_{i+1/2,j,k}).
\end{align}
By repeating the same procedure in the $y$ and $z$ directions, one is ready to
update the cell-centered conservative variables $\mathbf{U}_{i,j,k}$.

In computing the numerical flux, we also calculate the appropriate averages of the
transverse electromagnetic components at each face, and store them on a
working array. They will be used later to update the face-centered
electromagnetic field as discussed in the next subsection. One may adopt the
HLL average for this purpose \citep[e.g.,][]{2007A&A...473...11D,2015JCP},
which again reduces to the simple arithmetic average due to the nature of the
wave speed in this system. For instance, the $z$ component of the electromagnetic
fields at the $x$ and $y$ faces are calculated as follows:
\begin{align}
 &E_{z;i+1/2,j,k} \equiv
 \frac{1}{2} \left(
 {\mathcal R}^{\rmL_x} \left( E_{z;i,j,k} \right) +
 {\mathcal R}^{\rmR_x} \left( E_{z;i,j,k} \right)
 \right)
 \label{eq:hll_avg_ez_x}
 \\
 &E_{z;i,j+1/2,k} \equiv
 \frac{1}{2} \left(
 {\mathcal R}^{\rmL_y} \left( E_{z;i,j,k} \right) +
 {\mathcal R}^{\rmR_y} \left( E_{z;i,j,k} \right)
 \right)
 \label{eq:hll_avg_ez_y}
 \\
 &B_{z;i+1/2,j,k} \equiv
 \frac{1}{2} \left(
 {\mathcal R}^{\rmL_x} \left( B_{z;i,j,k} \right) +
 {\mathcal R}^{\rmR_x} \left( B_{z;i,j,k} \right)
 \right)
 \label{eq:hll_avg_bz_x}
 \\
 &B_{z;i,j+1/2,k} \equiv
 \frac{1}{2} \left(
 {\mathcal R}^{\rmL_y} \left( B_{z;i,j,k} \right) +
 {\mathcal R}^{\rmR_y} \left( B_{z;i,j,k} \right)
 \right)
 \label{eq:hll_avg_bz_y}
\end{align}
By repeating the same procedure to obtain the transverse quantities at each
face, all the electromagnetic field components are defined at each face.

\subsection{Update for face-centered variables}
A CT-type scheme requires the numerical flux defined at the edge center. More
specifically, $\hat{E}_{x;i,j+1/2,k+1/2}$, $\hat{E}_{y;i+1/2,j,k+1/2}$,
$\hat{E}_{z;i+1/2,j+1/2,k}$ are needed to update the magnetic field without
violating the divergence-free condition. Because it is defined at the edge
center, the flux definition must involve four states, contrary to the 1D
Riemann problem defined at the face, which involves only two states. This
clearly indicates that the Riemann problem is genuinely 2D in nature, but not
3D because the normal component is not involved
\citep{2004JCoPh.195...17L,2005JCoPh.205..509G,2010JCoPh.229.1970B}. Therefore,
use of an appropriate multidimensional Riemann solver is desired.

Now we consider evaluation of the flux defined at the $z$ edge:
$\hat{E}_{z;i+1/2,j+1/2,k}$, $\hat{B}_{z;i+1/2,j+1/2,k}$. The flux must be
calculated by using a 2D Riemann problem at the edge center specified by the
four states: $L_x L_x$, $L_x R_y$, $R_x L_y$, $R_x R_y$. Although the problem
is in general very difficult to solve, it is possible to obtain a particularly
simple expression with a 2D extension of HLL Riemann solver
\citep{Kurganov:2001:SCS:587161.587347,2004JCoPh.195...17L}. Here, the
constant maximum phase speed once again substantially simplifies the
expression. Consequently, we get
\begin{align}
 & \hat{E}_{z;i+1/2,j+1/2,k} =
 \frac{
 E_{z}^{\rmL_x \rmL_y} + E_{z}^{\rmL_x \rmR_y} +
 E_{z}^{\rmR_x \rmL_y} + E_{z}^{\rmR_x \rmR_y}
 }{4}
 \nonumber \\
 &\quad
 - \frac{B_x^{\rmR_y} - B_x^{\rmL_y}}{2}
 + \frac{B_y^{\rmR_x} - B_y^{\rmL_x}}{2},
 \label{eq:hll_ez_uct}
 \\
 & \hat{B}_{z;i+1/2,j+1/2,k} =
 \frac{
 B_{z}^{\rmL_x \rmL_y} + B_{z}^{\rmL_x \rmR_y} +
 B_{z}^{\rmR_x \rmL_y} + B_{z}^{\rmR_x \rmR_y}
 }{4}
 \nonumber \\
 &\quad
 + \frac{E_x^{\rmR_y} - E_x^{\rmL_y}}{2}
 - \frac{E_y^{\rmR_x} - E_y^{\rmL_x}}{2}.
 \label{eq:hll_bz_uct}
\end{align}
The other fluxes $\hat{E}_{x;i,j+1/2,k+1/2}$, $\hat{B}_{x;i,j+1/2,k+1/2}$,
$\hat{E}_{y;i+1/2,j,k+1/2}$, $\hat{B}_{y;i+1/2,j,k+1/2}$ are similarly
obtained by the cyclic permutation of indices. Notice that the above flux formula
automatically and correctly reduces to the 1D flux formula when homogeneity in
one direction is assumed. The second and third terms play the role for
upwinding, which were ignored in earlier attempts to combine a Riemann
solver with the CT-type discretization
\citep[e.g.,][]{1998ApJ...509..244R,1998JCoPh.142..331D,1999JCoPh.149..270B}.

In numerical implementation, we use a simplified approach rather than directly
obtaining the four states needed for the calculation of the flux
\citep[][]{2015JCP}. Because we already have all the electromagnetic
field components defined at each face, we can apply 1D reconstruction again to
estimate the four-point average in Eqs.~(\ref{eq:hll_ez_uct}) and
(\ref{eq:hll_bz_uct}). For instance, the numerical fluxes
$\hat{E}_{z;i+1/2,j+1/2,k}$ and $\hat{B}_{z;i+1/2,j+1/2,k}$ may be given as
follows
\begin{align}
 &
 \scalebox{0.8}{$\displaystyle
 \hat{E}_{z;i+1/2,j+1/2,k} =
 $}
 \nonumber \\
 &
 \scalebox{0.8}{$\displaystyle
 \frac{
 E^{\rmL_y}_{z;i+1/2,j+1/2,k} +
 E^{\rmR_y}_{z;i+1/2,j+1/2,k} +
 E^{\rmL_x}_{z;i+1/2,j+1/2,k} +
 E^{\rmR_x}_{z;i+1/2,j+1/2,k}
 }{4}
 $}
 \nonumber \\
 - &
 \scalebox{0.8}{$\displaystyle
 \frac{
 B^{\rmR_y}_{x;i+1/2,j+1/2,k} -
 B^{\rmL_y}_{x;i+1/2,j+1/2,k}
 }{2}
 $}
 \nonumber \\
 + &
 \scalebox{0.8}{$\displaystyle
 \frac{
 B^{\rmR_x}_{y;i+1/2,j+1/2,k} -
 B^{\rmL_x}_{y;i+1/2,j+1/2,k}
 } {2}
 $}
 \label{eq:hll_ez_flx}
 \\
 &
 \scalebox{0.8}{$\displaystyle
 \hat{B}_{z;i+1/2,j+1/2,k} =
 $}
 \nonumber \\
 &
 \scalebox{0.8}{$\displaystyle
 \frac{
 B^{\rmL_y}_{z;i+1/2,j+1/2,k} +
 B^{\rmR_y}_{z;i+1/2,j+1/2,k} +
 B^{\rmL_x}_{z;i+1/2,j+1/2,k} +
 B^{\rmR_x}_{z;i+1/2,j+1/2,k}
 }{4}
 $}
 \nonumber \\
 + &
 \scalebox{0.8}{$\displaystyle
 \frac{
 E^{\rmR_y}_{x;i+1/2,j+1/2,k} -
 E^{\rmL_y}_{x;i+1/2,j+1/2,k}
 }{2}
 $}
 \nonumber \\
 - &
 \scalebox{0.8}{$\displaystyle
 \frac{
 E^{\rmR_x}_{y;i+1/2,j+1/2,k} -
 E^{\rmL_x}_{y;i+1/2,j+1/2,k}
 }{2}
 $}
 \label{eq:hll_bz_flx}
\end{align}
Notice that the first four terms of these equations represent the arithmetic
mean of successive 1D reconstruction-averaging procedures. For instance, the
first two terms in Eq.~(\ref{eq:hll_ez_flx}) give the reconstruction-averaging
in the $y$ direction for $E_{z,i+1/2,j,k}$, which itself is a result of the
same procedure applied in the $x$ direction.  Similarly, the third and fourth
terms are results of the same procedure in the $y$ direction followed by the
$x$ direction. The fifth and sixth terms are obtained by the same
reconstruction procedure, but for the primary variable $B_{x;i+1/2,j,k}$
defined at the $x$ face. The same applies to the last two terms for
$B_{y;i,j+1/2,k}$ defined at the $y$ face. These numerical fluxes reduce to
the original definition Eqs.~(\ref{eq:hll_ez_uct}) and (\ref{eq:hll_bz_uct})
for the first-order piecewise constant reconstruction. This is true
even at higher orders, within the accuracy of reconstruction itself. We indeed
demonstrate that the second order accuracy was achieved in the numerical
experiments shown in section 4. It is important that the above numerical flux
formula yet retains the upwind property, and the definition of the flux
reduces to that of 1D HLL flux in the 1D limit.

The flux formula used in this study can be recognized as the simplest kind of
approximate Riemann solver in 2D. Although this approach has been successful
\citep{2004JCoPh.195...17L,2007A&A...473...11D,2014JCoPh.275..197A,
2015ApJ...808...54M}, the HLL Riemann solver is known to suffer excessive
numerical dissipation. In principle, this drawback can be overcome by using
more advanced multidimensional Riemann solvers that take into account variation
within the Riemann fan. Such multidimensional Riemann solvers have recently
been proposed by \cite{2010JCoPh.229.1970B,2012JCoPh.231.7476B}. However,
again the eigenstructure must be known in advance, to take advantage of those
sophisticated techniques.

In the absence of the charge and current densities, the above numerical flux
advance the electromagnetic field without violating the divergence-free
property. In other words, the divergence-free part evolves in a fully
consistent fashion. For the electric field, there exists a finite curl-free
part in the presence of non-zero charge density. The curl-free part
calculated from time integration of Ampere's law must be consistent with the
constraint Eq.~(\ref{eq:dive}). This can easily be confirmed by taking the
divergence of Ampere's law:
\begin{align}
 \frac{\partial}{\partial t} \div \mathbf{E}
 &= \div \left( c \rot \mathbf{B} - \mathbf{J} \right)
 \nonumber \\
 &=-\div \mathbf{J} = \frac{\partial}{\partial t} \varrho,
\end{align}
where the last equality comes from the charge conservation law. Namely, once
Gauss's law $\div \mathbf{E} = \varrho$ is satisfied at the initial condition,
it must be so at all times. A discrete analog of this relationship must be
satisfied for consistency in the time evolution.

Because we use the above mentioned flux formula, the condition $\div \left(
\rot \mathbf{B} \right)$ is satisfied at the discrete level. It is thus sufficient
to consider a proper choice of the charge and current densities. Recall that
the sixth component of the conservation law Eq.~(\ref{eq:mass_dif}) actually
gives the charge continuity equation. Therefore, the sixth component of the
fluxes $\mathbf{\hat{F}}_{i+1/2,j,k}$, $\mathbf{\hat{G}}_{i,j+1/2,k}$,
$\mathbf{\hat{H}}_{i,j,k+1/2}$ may respectively be identified as
$J_{x;i+1/2,j,k}$, $J_{y;i,j+1/2,k}$, $J_{z;i,j,k+1/2}$. By using the current
density defined by fluxes for the charge continuity equation to update the
electric field, one may guarantee that Gauss's law is always satisfied. This
is indeed the reason why we have designed the mesh such that that the current
density defined above and the primary electric field are collocated at the
same positions.

We now explicitly prove this is indeed the case. For simplicity, we here
consider a 2D case in the $x{\rm -}y$ plane and omit the index $k$ for the
third dimension, but the extension to 3D is straightforward. Ampere's law for
$E_x$ and $E_y$ may be written in the semidiscrete form as
\begin{align}
 &\frac{d}{dt} E_{x;i+1/2,j} =
 + \frac{c}{\Delta y}
 \left( \hat{B}_{z;i+1/2,j+1/2} - \hat{B}_{z;i+1/2,j-1/2} \right)
 \nonumber \\
 &\quad
 - J_{x;i+1/2,j},
 \label{eq:discrete-ampere-x} \\
 &\frac{d}{dt} E_{y;i,j+1/2} =
 - \frac{c}{\Delta x}
 \left( \hat{B}_{z;i+1/2,j+1/2} - \hat{B}_{z;i-1/2,j+1/2} \right)
 \nonumber \\
 &\quad
 - J_{y;i,j+1/2}.
 \label{eq:discrete-ampere-y}
\end{align}
From the finite difference approximation for $\partial E_x/\partial x$ and
$\partial E_y/\partial y$, we have
\begin{align}
 \frac{d}{dt}
 &
 \left(
 \frac{E_{x;i+1/2,j} - E_{x;i-1/2,j}}{\Delta x}
 \right)
 =
 \nonumber \\
 &-
 \left(
 \frac{J_{x;i+1/2,j} - J_{x;i-1/2,j}}{\Delta x}
 \right)
 \nonumber \\
 + \frac{c}{\Delta x \Delta y}
 &
 \left(
 \hat{B}_{z;i+1/2,j+1/2} - \hat{B}_{z;i+1/2,j-1/2} -
 \right.
 \nonumber \\
 &
 \left.
 \hat{B}_{z;i-1/2,j+1/2} + \hat{B}_{z;i-1/2,j-1/2}
 \right)
 \\
 \frac{d}{dt}
 &
 \left(
 \frac{E_{y;i,j+1/2} - E_{y;i,j-1/2}}{\Delta y}
 \right)
 =
 \nonumber \\
 &-
 \left(
 \frac{J_{y;i,j+1/2} - J_{x;i,j-1/2}}{\Delta x}
 \right)
 \nonumber \\
 - \frac{c}{\Delta x \Delta y}
 &
 \left(
 \hat{B}_{z;i+1/2,j+1/2} - \hat{B}_{z;i-1/2,j+1/2} -
 \right.
 \nonumber \\
 &
 \left.
 \hat{B}_{z;i+1/2,j-1/2} + \hat{B}_{z;i-1/2,j-1/2}
 \right)
\end{align}
This gives the time derivative of discrete divergence
\begin{align}
 \frac{d}{dt}
 &
 \left(
 \frac{E_{x;i+1/2,j} - E_{x;i-1/2,j}}{\Delta x} +
 \frac{E_{y;i,j+1/2} - E_{y;i,j-1/2}}{\Delta y}
 \right)
 \nonumber \\
 & =
 - \left(
 \frac{J_{x;i+1/2,j} - J_{x;i-1/2,j}}{\Delta x} +
 \frac{J_{y;i,j+1/2} - J_{y;i,j-1/2}}{\Delta y}
 \right)
 \nonumber \\
 & = \frac{d}{dt} \varrho_{i,j}.
\end{align}
Note that terms associated with the magnetic field have canceled out as a
result of the CT-type discretization, and the last equality comes from the
sixth component of the conservation law Eq.~(\ref{eq:fluid}). This proves that
the numerical solution always satisfies Gauss's law provided that it does so
at the initial condition.

\subsection{Summary of numerical procedure}
Here we summarize the numerical procedure used in this work. We
initialize the primitive variables at appropriate collocation points, i.e.,
the two-fluid quantities at cell centers, and the electromagnetic field at
edge centers. The initial condition must satisfy the two divergence
constraints. The primary electromagnetic field is interpolated to cell centers
and then the primitive variables defined at cell centers are converted to the
fluid conservative variables. This completes the preparation for time
integration, for which the third-order TVD Runge-Kutta scheme is used
throughout in this paper \citep{1988JCoPh..77..439S}.

Our numerical procedure for each substep of the Runge-Kutta integration is
summarized as follows:
\begin{enumerate}
 \item Reconstruction of primitive variables defined at cell centers is
       performed in each direction to estimate left and right states at each
       face.
 \item Numerical fluxes for the cell-centered fluid conservative variables are
       calculated using the 1D HLL flux formula Eq.~(\ref{eq:hll}). At the
       same time, the transverse electromagnetic field is calculated at each
       face by the HLL average formulae
       Eqs.~(\ref{eq:hll_avg_ez_x})-(\ref{eq:hll_avg_bz_y}). These
       face-centered transverse components are stored on a working array for
       later use.
 \item Reconstruction of the electromagnetic field defined at face centers
       (i.e., primary normal component and HLL-averaged transverse components)
       is performed. Numerical fluxes for Maxwell's equations are calculated
       using the 2D HLL flux formulae
       Eqs.~(\ref{eq:hll_ez_flx}) and ~(\ref{eq:hll_bz_flx}).
 \item The fluid conservative variables are updated using the face-centered
       flux obtained by Eq.~(\ref{eq:hll}). The contribution of the source
       terms $\mathbf{Q}$ on the right-hand side is also calculated using the
       cell-centered quantities. Similarly, the face-centered primary
       electromagnetic field components are updated using the edge-centered
       flux calculated by Eqs.~(\ref{eq:hll_ez_flx}) and
       (\ref{eq:hll_bz_flx}). The sixth component of face-centered flux for
       each direction is used as the current density to update the electric
       field using Ampere's law Eqs.~(\ref{eq:discrete-ampere-x}) and
       (\ref{eq:discrete-ampere-y}).
 \item The updated primary electromagnetic field is interpolated to cell
       centers. From the updated cell-centered quantities, the primitive
       variables are finally obtained using the primitive recovery algorithm
       described in section 3.1.
\end{enumerate}

\section{Numerical Results}
\label{result}

We here present numerical results for several test problems obtained with our
new code. Our primary concerns are overall accuracy of the scheme and
shock-capturing capability, or in other words, suppression of spurious
oscillations near discontinuities. In addition, the two-fluid effect must
be appropriately described if the resolution is sufficient, otherwise it
should not deteriorate the performance so that the RMHD result is
reproduced. We also present test problems where finite resistivity plays the
role. We have confirmed that the results satisfy the divergence constraints up
to machine precision in multidimensional problems, as is consistent with the
design of the scheme. In the following, we always set $c = 1$. A polytropic
index of $\Gamma = 4/3$ and resistivity of $\eta = 0$ were used unless
otherwise noted.

For problems that were originally proposed for RMHD (subsections 4.2,
4.3, 4.4, 4.5), one has to carefully consider the initialization of two-fluid
quantities. We always assumed that the charge neutrality is satisfied in the
initial condition. Therefore, the mass densities for proton (or positrons) and
electrons were given by
\begin{align}
 \rho_p = \frac{m_p}{m_p + m_e} \rho, \quad
 \rho_e = \frac{m_e}{m_p + m_e} \rho,
\end{align}
where $\rho$ is the total mass density (or MHD density). Similarly, the
temperatures for the two species were taken to be equal. The pressures for
each species were thus given by $p_p = p_e = p/2$, where $p$ is the total gas
pressure (or MHD gas pressure). The center-of-mass velocity was assumed to be
the RMHD four velocity. In applying the RTFED model to a RMHD problem, the
charge-to-mass ratios for protons (or positrons) $\mu_p$ and the ion to
electron mass ratio $m_p/m_e$ are free parameters that can be chosen, in
principle, arbitrarily. (The charge-to-mass ratio for electrons can be written
as $\mu_e = - \mu_p m_p/m_e$.) The total plasma skin depth is given in terms
of $\mu_p$ by
\begin{align}
 \lambda_p = \frac{c}{\omega_p} =
 \frac{c}{\mu_p} \sqrt{\frac{m_e}{m_p} \frac{1}{\rho}}.
\end{align}
Therefore, by appropriately choosing $\mu_p$, we can control the scale length
at which the two-fluid effect becomes apparent. Note that the effective skin
depth in a relativistically hot plasma becomes longer than $\lambda_p$ due to
relativistic inertia increase. Also, the above definition roughly corresponds
to the electron skin depth in an electron-proton plasma. The proton skin depth
is $\sqrt{m_p/m_e}$ times larger, and accordingly the proton inertia effect
(or the Hall effect) appears first on a larger scale.

\subsection{Circularly polarized wave}
We have checked the accuracy of our code for a smooth profile by using
circularly polarized (CP) waves in 2D. Propagation of a finite amplitude CP
\Alfven wave along the ambient magnetic field has commonly been adopted for
testing the accuracy of non-relativistic MHD codes. This is because such a
wave gives the exact solution even for an arbitrarily large amplitude, and the
numerical solution can be directly compared with the analytic
prediction. \cite{2007A&A...473...11D} had extended the analytic solution to
the relativistic regime, with which the accuracy of RMHD codes have been measured
in recent studies. We here need further generalization to the RTFED
equations. Although this involves a numerical solution of a somewhat complicated
dispersion relation, this is an ideal test to measure the accuracy of the
code. The detailed description of the dispersion relation is given in Appendix
\ref{circular}.

In a coordinate system specified by orthogonal unit vectors $\mathbf{e}_{i}
(i=1,2,3)$ with $\mathbf{e}_1$ being along the ambient magnetic field, the
exact solution of the wave propagating along $x_1$ can be written as
\begin{align}
 &
 B_1 = B_0, \quad
 B_2 = \xi B_0 \cos \phi, \quad
 B_3 =-\xi B_0 \sin \phi,
 \\
 &
 E_1 = 0, \quad
 E_2 =-\frac{\omega}{k c} \xi B_0 \sin \phi, \quad
 E_3 =-\frac{\omega}{k c} \xi B_0 \cos \phi,
 \\
 &
 u_{s,1} = 0, \quad
 u_{s,2} = U_{s} \cos \phi, \quad
 u_{s,3} =-U_{s} \sin \phi,
\end{align}
where $\phi = k x_1 - \omega t$ is the wave phase. Because it is an
incompressive mode, the density, and pressure are constant. However, note that
for a finite amplitude wave, the proper number density between the
two fluids is different because of the charge neutrality assumption. The
amplitude of velocity is determined by
\begin{align}
 U_s = - \xi \frac{\bar{\Omega}_{c,s}}{\omega + \bar{\Omega}_{c,s}/\gamma_{s}}
 \frac{\omega}{k},
\end{align}
where $\bar{\Omega}_{c,s} = \Omega_{c,s} / h_s$ is the cyclotron frequency
including the correction due to an effective inertia increase. As is discussed in
Appendix \ref{circular}, the dispersion relation is completely specified by
($\omega, k, \gamma_{p}, \gamma_{e}$). The dependence on the Lorentz factor
comes from a relativistic inertia increase corresponding to the quiver
motion. Therefore, for a given wavenumber $k$ and a normalized amplitude
$\xi$, one has to find a set of parameters $(\omega, \gamma_{p}, \gamma_{e})$
by solving the three-coupled nonlinear equations: (\ref{eq:cpw_dispersion})
and (\ref{eq:cpw_gamma}) for each fluid.

There is one subtlety in this test problem: a finite amplitude CP wave is
subject to a parametric instability. To prevent the growth of this physical
instability during the simulation time, one has to keep the wave amplitude
small enough. For this purpose, we used $\xi = 0.01$, which was found to be
sufficiently small to ignore the contamination for our second-order scheme.

We used a 2D rectangular simulation box, where the domain size in the $x$
direction ($2L$) is always twice as large as that in the $y$ direction
($L$). We used the same grid size for each direction, so that the number of
grid points were $2N$ and $N$ in $x$ and $y$ directions, respectively. The
ambient magnetic field was taken to be along the diagonal of the
mesh. Therefore, to set up the simulation, the analytic solution in the
orthogonal system $\mathbf{e}_i (i=1,2,3)$ was rotated by an angle $\theta =
\tan^{-1}(1/2)$ around $\mathbf{e}_3 = \mathbf{e}_z$ such that $\mathbf{e}_1$
points along the ambient magnetic field, whereas $\mathbf{e}_2$ is contained
in the $x{\rm -}y$ plane. The wavenumber was taken to be $(k_x, k_y) = (\pi/L,
2\pi/L)$, i.e., the wavelength was equal to the box sizes in each
direction. In this section, we only consider a pair plasma $m_p = m_e$, and an
effective magnetization parameter $\bar{\sigma}_e \equiv \sigma_e/h_e = 1$ was
used. The lab frame density for both species was also unity $\rho_p \gamma_p =
\rho_e \gamma_e = 1$, which ensures the charge neutrality condition. The
positron and electron temperatures were assumed to be equal $T_p/m_p c^2 =
T_e/m_e c^2 = 10^{-2}$, which gives $h_p = h_e = 1.04$. We used a
charge-to-mass ratio $\mu_p = -\mu_e = \sqrt{h_e}$, and a background magnetic
field $B_0 = c \sqrt{h_e \sigma_e}$. This normalizes the length and time to
the effective electron skin depth $\sqrt{h_e} c / \omega_p$, and the inverse
effective electron cyclotron frequency $\bar{\Omega}_{c,e}^{-1} = (\mu_e B_0 /
h_e c)^{-1}$, respectively. Notice that we have included the finite
temperature correction factor $h_e$ in the normalization. A finite wave
amplitude $\gamma_s \neq 1$ makes the actual skin depth different from the
unperturbed skin depth. Nevertheless, it remained small because $\gamma_s - 1
\lesssim 10^{-2}$ was satisfied for parameters adopted here.

The characteristics of the exact solution can be chosen by first specifying
the box size $L$ in unit of the skin depth. The system allows two distinct
solutions for a given $k$, one with a subluminal ($\omega/k < c$) and another
with a superluminal ($\omega/k > c$) phase speed. The simulations were
performed for two different box sizes $L = 64\pi, 4\pi$. Both superluminal and
subluminal modes were tested for each case. Consequently, there were four
cases in total. For each of these four cases, the parameters ($\omega$,
$\gamma_p$, $\gamma_e$) obtained by numerically solving the dispersion
relation with a tolerance of $10^{-12}$ are cataloged in
Table~\ref{table:parameter}. For $L = 64\pi$, because the wavelength is much
longer than the skin depth, the subluminal mode is essentially the \Alfven
wave in the RMHD regime (Case 1). The superluminal mode is an electromagnetic
wave with frequency close to the cut-off frequency (Case 2). On the other
hand, the runs with $L = 4\pi$ are in the regime where dispersive effect
becomes important. The subluminal and superluminal waves respectively
correspond to Case 3 and 4 in Table~\ref{table:parameter}. Except for Case 2,
the time step was chosen to be $\Delta t = (2\pi/\omega)/M$ where $M$ is the
minimum integer such that a CFL number is less than 0.25 is satisfied. For
Case 2, the condition that a CFL number is less than 0.10 was used instead to
ensure that the time step was small enough to resolve the wave frequency in
low-resolution runs.

\begin{table*}
 \begin{center}
  \caption{Parameters used for circularly polarized wave test problems.}
  \label{table:parameter}
  \begin{tabular}[t]{ccccc}
   \tableline
   Case & L & $\omega$ & $\gamma_p-1$ & $\gamma_e-1$
   \\ \tableline
   %%% parameters for problem 1
%   1 & $64 \pi$
%     & 4.03214677454124e-02
%     & 1.53832112446128e-05
%     & 1.80772971893894e-05
   1 & $64 \pi$
     & 4.03214677454 $\times 10^{-2}$
     & 1.53832112446 $\times 10^{-5}$
     & 1.80772971893 $\times 10^{-5}$
   \\
   %%% parameters for problem 2
%   2 & $64 \pi$
%     & 1.67851327468321e+00
%     & 4.02522132038374e-03
%     & 5.29202636700660e-02
   2 & $64 \pi$
     & 1.67851327468 $\times 10^{+0}$
     & 4.02522132038 $\times 10^{-3}$
     & 5.29202636700 $\times 10^{-2}$
   \\
   %%% parameters for problem 3
%   3 & $4 \pi$
%     & 5.63803828148330e-01
%     & 5.19940020571318e-06
%     & 6.68453076522190e-05
   3 & $4 \pi$
     & 5.63803828148 $\times 10^{-1}$
     & 5.19940020571 $\times 10^{-6}$
     & 6.68453076522 $\times 10^{-5}$
   \\
   %%% parameters for problem 4
%   4 & $4 \pi$
%     & 1.98281630723844e+00
%     & 1.76755626295488e-05
%     & 1.62742550300576e-04
   4 & $4 \pi$
     & 1.98281630723 $\times 10^{+0}$
     & 1.76755626295 $\times 10^{-5}$
     & 1.62742550300 $\times 10^{-4}$
   \\ \tableline
  \end{tabular}
 \end{center}
\end{table*}

For each of the four test cases, we have measured the error convergence by
changing the resolution in the range $N = 16, 32, 64, 128$, which is
summarized in Table~\ref{table:convergence}. The error was evaluated with both
$L_1$ and $L_{\infty}$ norms by the deviation from the initial condition after
five wave propagation periods. The errors in $E_y$ and $B_y$ components showed
a similar tendency. Therefore, only the errors in $E_y$ are shown in
Table~\ref{table:convergence}. We see that, except for Case 2, the code
roughly reproduced the second-order convergence consistent with the design
accuracy. The convergence in Case 2 was actually better than second order,
probably because the wave was highly superluminal, which means it is a
non-propagating mode. In this case, the accuracy in space is not important and
the error may be dominated by that in the time integration scheme. Because we
used the third-order TVD Runge-Kutta scheme for time integration, the error
convergence appeared to be closer to third order. In any case, we have
confirmed that the scheme achieved at least second-order overall accuracy.

\begin{table*}
 \begin{center}
  \caption{Numerical convergence for circularly polarized wave test
  problems. The first column corresponds to the Case ID in
  Table~\ref{table:parameter}.}

  \label{table:convergence}
  \begin{tabular}[t]{cccccc}
   \tableline
   Case
   & Number of mesh
   & $L_1$ error
   & $L_1$ order
   & $L_{\infty}$ error
   & $L_{\infty}$ order
   \\ \tableline
   %%% result for problem 1
   % 32     2.41646e-03  0.00     4.00605e-03  0.00
   % 64     4.97612e-04  2.28     9.22089e-04  2.12
   %128     1.26208e-04  1.98     2.13817e-04  2.11
   %256     3.19985e-05  1.98     5.28181e-05  2.02
   1
   & $32 \times 16$
   & 2.41646 $\times 10^{-3}$  & ---  & 4.00605 $\times 10^{-3}$ & ---
   \\
   & $64 \times 32$
   & 4.97612 $\times 10^{-4}$  & 2.28 & 9.22089 $\times 10^{-4}$ & 2.12
   \\
   & $128 \times 64$
   & 1.26208 $\times 10^{-5}$  & 1.98 & 2.13817 $\times 10^{-4}$ & 2.11
   \\
   & $256 \times 128$
   & 3.19985 $\times 10^{-5}$  & 1.98 & 5.28181 $\times 10^{-5}$ & 2.02
   \\ \tableline
   %%% result for problem 2
   % 32     7.41275e-02  0.00     1.16316e-01  0.00
   % 64     1.37123e-02  2.43     2.19120e-02  2.41
   %128     2.05654e-03  2.74     3.35127e-03  2.71
   %256     3.10960e-04  2.73     5.15282e-04  2.70
   2
   & $32 \times 16$
   & 7.41275 $\times 10^{-2}$ & ---  & 1.16316 $\times 10^{-1}$ & ---
   \\
   & $64 \times 32$
   & 1.37123 $\times 10^{-2}$ & 2.43 & 2.19120 $\times 10^{-2}$ & 2.41
   \\
   & $128 \times 64$
   & 2.05654 $\times 10^{-3}$ & 2.74 & 3.35127 $\times 10^{-3}$ & 2.71
   \\
   & $256 \times 128$
   & 3.10960 $\times 10^{-4}$ & 2.73 & 5.15282 $\times 10^{-4}$ & 2.70
   \\ \tableline
   %%% result for problem 3
   % 32     2.04812e-03  0.00     3.11209e-03  0.00
   % 64     3.66271e-04  2.48     5.91428e-04  2.40
   %128     7.87039e-05  2.22     1.25621e-04  2.24
   %256     1.99544e-05  1.98     3.13751e-05  2.00
   3
   & $32 \times 16$
   & 2.04812 $\times 10^{-3}$ & ---  & 3.11209 $\times 10^{-3}$ & ---
   \\
   & $64 \times 32$
   & 3.66271 $\times 10^{-4}$ & 2.48 & 5.91428 $\times 10^{-4}$ & 2.40
   \\
   & $128 \times 64$
   & 7.87039 $\times 10^{-5}$ & 2.22 & 1.25621 $\times 10^{-4}$ & 2.24
   \\
   & $256 \times 128$
   & 1.99544 $\times 10^{-5}$ & 1.98 & 3.13751 $\times 10^{-5}$ & 2.00
   \\ \tableline
   %%% result for problem 4
   % 32     3.19915e-03  0.00     5.24118e-03  0.00
   % 64     5.37798e-04  2.57     9.00981e-04  2.54
   %128     1.09563e-04  2.30     1.85969e-04  2.28
   %256     2.54872e-05  2.10     4.31465e-05  2.11
   4
   & $32 \times 16$
   & 3.19915 $\times 10^{-3}$ & ---  & 5.24118 $\times 10^{-3}$ & ---
   \\
   & $64 \times 32$
   & 5.37798 $\times 10^{-4}$ & 2.57 & 9.00981 $\times 10^{-4}$ & 2.54
   \\
   & $128 \times 64$
   & 1.09563 $\times 10^{-4}$ & 2.30 & 1.85969 $\times 10^{-4}$ & 2.28
   \\
   & $256 \times 128$
   & 2.54872 $\times 10^{-5}$ & 2.10 & 4.31465 $\times 10^{-5}$ & 2.11
   \\ \tableline
  \end{tabular}
 \end{center}
\end{table*}

\subsection{Generalized Brio-Wu problem}
The Brio-Wu shock tube problem is one of the standard test problems for
classical MHD. We here adopt a relativistic analog of the problem \citep[e.g.,
Balsara 2001,][]{2003A&A...400..397D}, which has been widely accepted in the
RMHD community. The RTFED model should be able to reproduce the RMHD result by
keeping the skin depth sufficiently small with respect to the grid size. On the
other hand, one expects that dispersive waves will appear if the resolution is
sufficient. This has also been shown in the original non-relativistic version
of the problem \citep[e.g.,][]{2006JCoPh.219..418H,2015JCP}.

A 1D simulation box of unit length ($0 \leq x \leq 1$) was initially divided
into the left and right states at the center $x = 0.5$. The left and right
states were given as follows
\begin{align}
 \begin{pmatrix}
  \rho \\
  p \\
  B_x \\
  B_y
 \end{pmatrix}_{\rm left}
 =
 \begin{pmatrix}
  1.0 \\
  1.0 \\
  0.5 \\
  1.0
 \end{pmatrix}
 , \quad
 \begin{pmatrix}
  \rho \\
  p \\
  B_x \\
  B_y
 \end{pmatrix}_{\rm right}
 =
 \begin{pmatrix}
  0.125 \\
  0.1 \\
  0.5 \\
  -1.0
 \end{pmatrix}.
\end{align}
Other quantities were initialized by zero. The problem was run with an
adiabatic index of $\Gamma = 2.0$. A CFL number of 0.1 was used.

\ifemulateapj
\FigureTwo
\fi

Fig.~\ref{fig:bw_mhd} shows the numerical solution at $t = 0.4$ for a pair
plasma with $\mu_p = -\mu_e = 10^{4}$.  The number of grid points was
1600. Because the skin depth $\lambda_p = 10^{-4} / \sqrt{\rho}$ was smaller
than the grid size in the entire box, the solution agreed quite well with the
RMHD result presented in the literature.

\ifemulateapj
\FigureThree
\FigureFour
\fi

On the other hand, Fig.~\ref{fig:bw_pair} and \ref{fig:bw_ep} show results for
$\mu_p = 10^{3}$ with $m_p/m_e = 1$ and $100$, respectively. The total density
profiles obtained with three different resolutions are shown in each panel: $N
= 400$ (red), $1600$ (green), $6400$ (blue). A reference solution
corresponding to the RMHD limit ($\mu_p = 10^{5}, m_p/m_e = 1, N = 6400$) is
also shown with a black line for comparison. The grid size in the lowest
resolution run was slightly larger than the skin depth. Thus, the numerical
solutions roughly agreed with the RMHD prediction, although discontinuities
were smeared out by numerical dissipation. As increasing the resolution,
dispersive waves due to the two-fluid effect clearly appeared in the
solutions. The results for a pair plasma and an electron-proton plasma are
qualitatively the same. The most noticeable difference is the dip in density
ahead (to the left) of the slow compound wave at $x \simeq 0.5$ in the
electron-proton case. A similar structure was also observed in the
non-relativistic case \citep{2006JCoPh.219..418H,2015JCP}.

\subsection{Orszag-Tang vortex}

The Orszag-Tang vortex problem has been used as a benchmark problem for
multidimensional MHD codes. Although the problem starts from a smooth profile,
the solution involves complicated multidimensional discontinuities. This
possibly produces non-negligible numerical error in the divergence-free
condition $\nabla \cdot \mathbf{B} = 0$, which may lead to collapse of the
numerical simulation. Here we adopt a relativistic analog of the problem to
demonstrate that our code is capable of describing complex multidimensional
flows involving discontinuities without numerical difficulty.

The initial condition was the same as that used in
\cite{2011ApJS..193....6B}. The simulation domain was a 2D unit square $0 \leq
x \leq 1, 0 \leq y \leq 1$ with the periodic boundary condition applied in
both directions. The initial condition for the density $\rho = \Gamma^2/4\pi$,
and pressure $p = \Gamma/4\pi$ were uniform, where we used $\Gamma = 5/3$. The
three velocity and the magnetic field were initialized as follows:
\begin{align}
 &
 V_x =-V_0 \sin\left(2 \pi y\right), \quad
 V_y = V_0 \cos\left(2 \pi x\right)
 \\
 &
 B_x =-B_0 \sin\left(2 \pi y\right), \quad
 B_y = B_0 \sin\left(4 \pi x\right),
\end{align}
where $V_0 = 1/2$ and $B_0 = 1/\sqrt{4 \pi}$, respectively. We assumed a pair
plasma $\mu_p = -\mu_e = 10^{3}$, which makes the skin depth smaller than the
grid size. However, the effective skin depth including the effect of
the relativistic temperature was found to be comparable to the grid size in the
numerical solution. Therefore, the result would be modified slightly from the
RMHD solution by the two-fluid effect. The out-of-plane magnetic field was zero
$B_z = 0$, whereas $V_{p,z} = - V_{e,z}$ was finite so as to satisfy Ampere's
law in the initial condition. The corresponding four velocity was given by
\begin{align}
 u_{p,z} = - u_{e,z} = \frac{4 \pi c B_0}{\mu_p \rho}
 \left(
 \cos \left(4 \pi x\right) + \frac{1}{2} \cos \left(2 \pi y\right)
 \right)
\end{align}
The in-plane components of the velocity for each fluid may be taken to be
equal: $V_{p,x} = V_{e,x}, V_{p,y} = V_{e,y}$. The electric field was
initialized by $\mathbf{E} = - \mathbf{V}/c \times \mathbf{B}$. Note that the
initial condition satisfies not only $\nabla \cdot \mathbf{B} = 0$, but also
$\nabla \cdot \mathbf{E} = 0$, the latter is consistent with the charge
neutrality $\mu_p \rho_p + \mu_e \rho_e = 0$.

The numerical solution with a $200 \times 200$ mesh at $t = 1$ is shown in
Fig.~\ref{fig:ot} for the density, gas pressure, magnetic pressure, and bulk
Lorentz factor, respectively. A CFL number of 0.2 was used. The density and
gas pressure agreed quite well with the published RMHD results
\citep{2011ApJS..193....6B}, whereas small scale features in the magnetic
pressure were slightly different due to the appearance of the two-fluid
effect.

\ifemulateapj
\FigureFive
\fi

\subsection{Strong cylindrical explosion}

The strong cylindrical explosion in a magnetized uniform medium has been a
stringent benchmark problem to test the robustness of a numerical scheme. We
adopt the relativistic version described in \cite{1999MNRAS.303..343K}. The
simulation box was a 2D square domain: $-6 \leq x \leq +6, -6 \leq y \leq
+6$. We used a $200 \times 200$ mesh. Initially, the density and pressure in
the central region $r < 0.8$ (where $r = \sqrt{x^2 + y^2}$ is the distance
from the origin) were $\rho_{in} = 10^{-2}$ and $p_{in} = 1.0$, which were
respectively higher than those in the uniform surrounding medium $\rho_{out} =
10^{-4}$, $p_{out} = 5 \times 10^{-4}$. Both the density and pressure linearly
decreased from the values of the inside ($\rho_{in}, p_{in}$) to the outside
($\rho_{out}, p_{out}$) in the range $0.8 \leq r \leq 1$. The plasma was
initially at rest and the electric field was zero. The initial magnetic field
was in the $x$ direction: $B_x = B_0$. The results of two runs with different
initial magnetic field strength $B_0$ are shown below. We considered a pair
plasma and the charge-to-mass ratio was chosen to be $\mu_p = -\mu_e =
10^{3}$. This gives a skin depth of $\lambda_p = 10^{-3} / \sqrt{\rho}$,
which is slightly larger than the grid size ($\Delta x = \Delta y = 0.06$) in
the surrounding uniform medium ($\lambda_p = 0.1$ for $\rho =
10^{-4}$). Simulations were run with a CFL number of 0.1.

\ifemulateapj
\FigureSix
\FigureSeven
\fi

Fig.~\ref{fig:blast_weak} shows the result for a weakly magnetized medium $B_0
= 0.1$. A strong shock expanded roughly symmetrically into the surrounding
medium. The numerical solution was quite similar to the RMHD result. The
maximum Lorentz factor was $\gamma_{\rm max} \simeq 3.95$. On the other hand,
Fig.~\ref{fig:blast_strong} shows the result for a strongly magnetized medium
$B_0 = 1.0$. In this case, the magnetic field in the surrounding medium was so
strong that the inner structure was significantly modified. As a result, the
expansion was primarily in the direction parallel to the ambient magnetic
field. The maximum Lorentz factor in this case reached $\gamma_{\rm max}
\simeq 3.00$. In general, this problem is known to be a stringent problem
for which many RMHD codes would fail unless ad hoc changes were introduced in
the code. Nevertheless, our code is able to keep track of the evolution of
the problem without any numerical tricks.

\subsection{Self-similar current sheet}

So far we have considered test problems without resistivity. The resistive
effect can be tested using the problem first presented in
\cite{2007MNRAS.382..995K}. We consider a 1D current sheet that involves
only variation in the $B_y$ component. When the magnetic pressure is much
smaller than the constant gas pressure, the evolution of the magnetic field in
a resistive medium may be approximated by the diffusion equation:
\begin{align}
 \frac{\partial B_y}{\partial t} -
 D \frac{\partial^2 B_y}{\partial x^2} = 0,
\end{align}
where the diffusion coefficient is given in terms of resistivity by $D = \eta
c^2$. A self-similar solution suggested by \cite{2007MNRAS.382..995K} is given
as follows
\begin{align}
 B_y (x,t) = B_0 \, {\rm erf} \left( \frac{x}{2 \sqrt{D t}} \right),
\end{align}
where ${\rm erf}(x)$ is the error function.

We set up the initial condition using the analytic solution at $t = 1$ with
$B_0 = 1$ and $\eta c^2 = D = 0.01$ in a 1D computational domain $-1.5 \leq x
\leq +1.5$ with 200 grid points. We used the conducting wall boundary
condition. The charge-to-mass ratios were taken as $\mu_p = -\mu_e =
10^{3}$. This gives the skin depth of $\lambda_p = 10^{-3} / \sqrt{\rho}$, which
is much smaller than the grid spacing $\Delta x = 1.5 \times 10^{-2}$. One can
thus expect that the RTFED equations should reproduce the RMHD result.

The $x$ and $y$ components of the magnetic field and the electric field were
zero. The density and pressure were uniform $\rho = 1$ and $p = 50$. The $x$
and $y$ components of velocity were zero, whereas the $z$ component was
initialized by
\begin{align}
 u_{p,z} = - u_{e,z} = \frac{c B_0}{\mu_p \rho \sqrt{\pi D}}
 \exp \left( - \frac{x^2}{4 D}\right).
\end{align}
This gives the conduction current that is consistent with Ampere's law.

Fig.~\ref{fig:selfsimilar} shows the numerical solution at $t = 9$ in parallel
with the analytic prediction. A CFL number of 0.5 was used for the
simulation. The two solutions were essentially indistinguishable. This indicates
that our choice of the friction term can reproduce the resistive RMHD result.

\ifemulateapj
\FigureEight
\fi

\subsection{Resistive perpendicular shock}
This test problem first proposed by \cite{2014MNRAS.438..704B} deals with a
fast magnetosonic shock propagating across the magnetic field in the two-fluid
regime. At a scale length comparable to the skin depth, a shock wave in the
two-fluid approximation in general involves a dispersive wave train either in
the upstream or downstream of the shock. For a perpendicular fast-mode shock,
the wave train appears only in the downstream because the group velocity $d
\omega/d k$ is a decreasing function of $k$. The amplitude of the wave train
gradually decreases due to dissipation as increasing the distance from the
shock. In the presence of a finite resistivity, one can check the convergence
of the numerical solution by increasing the resolution beyond the resistive
scale length.

\begin{table}
 \begin{center}
  \caption{Left and right states for resistive perpendicular shock problem.}
  \label{table:perpshock}
  \begin{tabular}[t]{ccc}
   \tableline
   Parameter & Left state & Right state
   \\
   \tableline
   $\rho$    & 1.0  & 2.059639
           \\
   $\gamma$  & 10.0 & 4.933298
           \\
   p         & 0.1  & 0.3420819
           \\
   $B_y$     & 60   & 60.9648752
           \\ \tableline
  \end{tabular}
 \end{center}
\end{table}

The simulation setup was essentially the same as that of the generalized
Brio-Wu problem discussed in section 4.2, except for different initial left
and right states. The initial left and right states were determined by solving
numerically the Rankine-Hugoniot relations for a perpendicular RMHD shock with
$\Gamma = 4/3$, and are given in Table \ref{table:perpshock}. The normal flow
velocity and the $z$ component of electric field were respectively given by
$u_x = \sqrt{1 + \gamma^2}$ and $E_z = - u_x B_y / \gamma c$, whereas other
components of vector quantities were initialized by zero. We considered a pair
plasma with a charge-to-mass ratio of $\mu_p = -\mu_e = 200$, which gives
$\lambda_p = 5 \times 10^{-3} / \sqrt{\rho}$. We adopted a constant
resistivity of $\eta = 2.5 \times 10^{-3}$, corresponding to a normalized
frictional relaxation time scale of $\tau_f = 1/\omega_p \eta =0.2$. The
magnetization parameter in the upstream (left) medium in this setup becomes
$\sigma_e = \sigma_p = 72$. Therefore, the problem deals with a strongly
magnetized plasma. This set of parameters is very similar to the case 1 of the
problem discussed in section 4.3 of \cite{2014MNRAS.438..704B}. However, our
setup was not exactly the same because we could not reproduce their results:
this is probably because of some errors in their description of the initial
condition.

\ifemulateapj
\FigureNine
\fi

In Fig.~\ref{fig:perpshock}, the density and magnetic field profiles at $t =
0.5$ are shown for three different resolutions: $N = 800, 1600, 3200$. A CFL
number of 0.2 was used in every run. Because the initial condition satisfied
the Ranking-Hugoniot relations, the shock structure was stationary in the
simulation frame. The shock transition did not involve a discontinuous
subshock but it did exhibit a laminar profile with a trailing wave train structure
as shown by \cite{2014MNRAS.438..704B}. It appeared that the solution with $N =
1600$ almost coincides with that of $N = 3200$, indicating that numerical
convergence was achieved at this resolution. In contrast, when the physical
resistivity was turned off, we did not observe the convergence of the
numerical solution. This is because ``resistive'' dissipation always occurs at
the grid scale. In any case, there was no numerical stability issue even
without finite resistivity.

\subsection{Magnetic reconnection}
Our final test problem is magnetic reconnection for a strongly magnetized
electron-proton plasma. Previous simulation studies of relativistic magnetic
reconnection in the two-fluid regime have been presented only for a pair
plasma. On the other hand, it is well known that magnetic reconnection becomes
efficient in a thin current sheet whose thickness is on the order of ion skin
depth in a non-relativistic electron-proton plasma. Here we demonstrate that
essentially the same argument applies to the relativistic regime and fast
magnetic reconnection is realized.

The simulation setup was an extension of a non-relativistic GEM (Geospace
Environment Modeling) magnetic reconnection problem
\citep[e.g.,][]{2001JGR...106.3715B}. The initial magnetic field and proper
number density profiles were given by
\begin{align}
 B_{x} = B_0 \tan \left(y / d\right), B_{y} = B_{z} = 0,
\end{align}
and
\begin{align}
 n = n_0 {\rm sech}^2 \left(y / d\right) + n_{\rm bg}
\end{align}
respectively. The number density was assumed to be the same between the
species. The system is characterized by the magnetization parameter
$\sigma_{s} = B_0^2/ n_0 m_s c^2$. It is readily seen that the Alfven speed is
approximately given by $v_A/c \simeq \sqrt{\sigma_p/(1 + \sigma_p)}$ for
$m_p/m_e \gg 1$. The relativistic effect thus becomes important for magnetic
reconnection in a strongly magnetized plasma $\sigma_p \gtrsim 1$. The initial
temperature was given by $T_s/m_s c^2 = \sigma_s / 4$, which is consistent with
the pressure balance condition for an equal temperature plasma $T_p =
T_e$. Similarly, we assumed that the initial current is carried equally by the
two species. This leads to the $z$ component of four velocity
\begin{align}
 u_{z,p} = -u_{z,e} = \frac{c}{2 e d}
 \frac{B_0 {\rm sech}^2 (y/d)}{n_0 {\rm sech}^2 (y/d) + n_{\rm bg}},
\end{align}
for consistency with Ampere's law. Other components of velocity were
initialized by zero. Notice that because the Lorenz factors are the same between
the species with this setup, the charge neutrality $\nabla \cdot \mathbf{E} =
0$ in the initial condition is automatically satisfied. We thus assumed that
the electric field was initially zero everywhere in the simulation box. In
addition, to initiate magnetic reconnection, the magnetic field was perturbed
by introducing the out-of-plane vector potential of the form
\begin{align}
 \Phi = \alpha B_0
 \cos \left( \frac{\pi x}{L} \right)
 \cos \left( \frac{\pi y}{L} \right),
\end{align}
where $L$ is the system size in $y$ direction and $\alpha = 0.1$ is the
amplitude of perturbation. More specifically, the two-dimensional
computational domain $-L \leq x \leq +L$, $-L/2 \leq y \leq +L/2$ was
used. The number of grids in $x$ and $y$ directions were $2N$ and $N$,
respectively. Thus, the grid sizes in each direction were the same $\Delta x =
\Delta y$.

We adopt a normalization such that $c = m_p = n_0 = 1$ and $B_0 = 1/e =
\sqrt{\sigma_p}$. With this normalization, time and length are measured in
units of $\Omega_{c,p}^{-1} = (e B_0 / m_p c)^{-1}$, and $c/\Omega_{c,p}$,
respectively. A system size of $L/c/\Omega_{c,p} = 12.8$, and a thickness of
the current sheet of $d/c/\Omega_{c,p} = 1$ were used. The boundary condition
in the $x$ direction was periodic, whereas the conducting wall boundary
condition was used in the $y$ direction. A time step of $\Omega_{c,p} \Delta t
= 10^{-3}$ was used for all the simulations shown below.

In the following, we show simulation results obtained with $m_p/m_e = 25$,
$\sigma_p = 1.0$, $\sigma_e = 25.0$. A background density of $n_{\rm bg}/n_0 =
0.2$ was used. In such a strongly magnetized current sheet, one has to take
into account the effective inertia increase due to relativistic
temperature. Consequently, the effective magnetization ratios become
$\bar{\sigma}_p = \sigma_p/h_p = 1/2$ and $\bar{\sigma}_e = \sigma_e/h_e =
25/26$, respectively. Therefore, the Alfven speed was roughly 57\% of the
speed of light. We used a constant resistivity of $\eta c^2 = 0.01$. Time
development of the normalized reconnected magnetic flux calculated by
\begin{align}
 \psi(t) = \frac{1}{2 B_0} \int_{-L}^{L} \| B_y(x,y=0,t) \| dx
\end{align}
is shown in Fig.~\ref{fig:recflux} for three runs with increasing resolutions:
$N = 128, 256, 512$. All three runs exhibited almost the same evolution. In
particular, the results are almost indistinguishable up to $\Omega_{c,p} t
\simeq 70$. In the late phase, the interaction between fast reconnection
outflows and a plasmoid produced complicated structures involving
discontinuities, which is probably the main reason for the slight
deviation. The reconnection rate defined as the inflow speed toward the
neutral sheet in units of Alfven speed $v_{\rm in}/v_A$ was estimated from the
slope of the reconnected flux. The peak reconnection rate reached as high as
$\sim 0.13$ at $\Omega_{c,p} t \simeq 62$. Similar reconnection rates are
typically observed in Hall-MHD simulations for a non-relativistic
electron-proton plasma.

\ifemulateapj
\FigureTen
\fi

Fig.~\ref{fig:mrx1} and \ref{fig:mrx2} show two snapshots at $\Omega_{c,p} t =
40, 80$ for the total mass density $\rho$, out-of-plane magnetic field $B_z$,
$x$ component of four velocity for protons $u_{p,x}$ and electrons $u_{e,x}$,
respectively. The reconnection occurred around the origin from which fast
bipolar outflows were ejected both in the positive and negative $x$ directions. In
Fig.\ref{fig:mrx1}, we can clearly see the quadrupolar out-of-plane magnetic field
$B_z$ around the X-point generated by the Hall effect. The outflow speed for
protons was slower than electrons, indicating the decoupling between the
species. At a later time, the outflow is accelerated even further. The ion
outflow speed reached as high as the Alfven speed, whereas the electron
outflow was accelerated essentially to the speed of light.

Although these qualitative features are essentially unchanged from the
non-relativistic counterpart, there are quantitative differences. For
instance, the observed magnitude of $B_z$ at early times was much smaller,
which is somehow to be expected in the relativistic regime. The electron fluid
had an increased effective inertia due to relativistic temperature $T_e/m_e
c^2 = 6.25$, whereas the same effect was less important for protons $T_p/m_p
c^2 = 0.5$. Consequently, the ratio of effective inertia between protons and
electrons was reduced to $m_p (1 + 4 T_p/m_p c^2) / m_e (1 + 4 T_e/m_e c^2)
\sim 1.9$ (where $\Gamma = 4/3$ was used). This made the decoupling of the
dynamics between the two fluids becomes relatively weak, resulting in the
reduced amplitude of Hall magnetic field. Therefore, in an extreme situation
$\sigma_p \gg m_p/m_e$, we expect that the dynamics will essentially become
the same as that in a pair plasma as long as the temperatures of the two
species remain the same. A more detailed study will be presented elsewhere in
the future.

\ifemulateapj
\FigureEleven
\FigureTwelve
\fi

\section{Conclusions}
\label{conclusion}

In this paper, we have discussed the relativistic two-fluid electrodynamics
(RTFED) equations, which are an extension of relativistic magnetohydrodynamics
(RMHD). The advantage of the RTFED model is obviously its capability for a
wider range of applications. In contrast to RMHD, there is no inherent
difficulty for dealing with a region where the local electric field is larger
than the magnetic field, which may become important for extreme environments
in high energy astrophysics. Also, it is easy to implement a finite
resistivity without suffering from the singularity at infinite
conductivity. The resistivity (or the friction term) introduced in this paper
reduces to the one used in current generation resistive RMHD codes in the long
wavelength limit. This fact remains valid regardless of the proton-to-electron
mass ratio, which makes it possible to investigate the resistive effect in not
only a pair plasma but also an electron-proton plasma.

A 3D simulation code solving the RTFED equations has been described. The code
achieves overall second-order accuracy for smooth profiles. If the grid size
is taken to be large compared to the skin depth, the RMHD
shocks/discontinuities are captured without appreciable numerical
oscillation. Furthermore, dispersive waves arising from the two-fluid effect
are correctly described in cases where sufficient resolution is available. The
numerical algorithm presented here guarantees that the two divergence
constraints for the electromagnetic field are preserved up to machine
precision.

It is also possible to extend the code to higher order. Indeed,
\cite{2016JCoPh.318..169B} have presented up to a fourth-order accurate finite
volume implementation for the RTFED equations. They have also invented a novel
and more consistent reconstruction scheme of the electric field that satisfies
Gauss' law over the entire control volume. Alternatively, one may also adopt a
finite difference approach, in which case a higher-order
reconstruction/interpolation can be applied in a dimension-by-dimension
manner. Note that, even in this case, it is possible to construct a scheme
that exactly preserves the divergence constraints
\citep[e.g.,][]{2007A&A...473...11D}.

To be fair, there is one critical disadvantage in the RTFED equation. Because
it includes high-frequency plasma waves as eigenmodes even in the long
wavelength limit, the numerical stability inevitably requires a small time
step to resolve the plasma frequency. This is the most serious obstacle for
the model when application to macroscopic phenomena (in the presence of a
dense plasma) is considered. In these situations, the dynamical time scale and
the inverse plasma frequency will differ by orders of magnitude. A naive way
to resolve the issue is to use an implicit time integration scheme
\citep[e.g.,][]{Kumar2012,2016JCoPh.318..169B}. Indeed, because the high-frequency
waves in the long wavelength limit are non-propagating, the scheme can be made
locally implicit. This is advantageous because it will not require communications
with neighboring processors in parallelization on a distributed memory
system. It may also be possible to utilize analytic solutions for such
high-frequency waves combined with the operator splitting technique if the
wave amplitude remains sufficiently small.

Another numerical issue, although less important than the one above, is associated
with the use of the simplified Riemann solver. Because the maximum phase speed
in the RTFED equations is always given by the speed of light, it may introduce
excessive numerical dissipation in situations where the RMHD characteristics
have only non-relativistic speeds. In a sense, if the dynamical time scale is
much less than the light transit time, one may think that the dynamics of the
plasma are decoupled from the electromagnetic wave propagation. In principle,
taking advantage of the decoupling, it would be possible to construct a more
sophisticated Riemann solver that resolves the internal structure of the
Riemann fan. This allows one to obtain higher accuracy both in
non-relativistic and relativistic regions at the same time.

Despite the numerical issues raised above, given the potential of the RTFED
equations, it is important to continue investigating of numerical methods
to overcome the weak points. This will certainly extend the applicability of
the model in the field of high energy astrophysics.

\acknowledgments

This work was supported by KAKENHI 25800101 from JSPS of Japan. The author is
indebted to D.~Balsara, and K.~Hirabayashi for stimulating discussion on
numerical methods, and to S.~Zenitani for his comments on the manuscript.

\appendix

\newpage
\section{Finite Amplitude Circularly Polarized Wave}
\label{circular}

It has been known that there exists an analytic solution to a finite amplitude
circularly polarized waves propagating along the ambient magnetic field for
the relativistic cold two-fluid plasma equations
\citep{1976JPlPh..15..335K}. Because it does not involve density perturbations,
it is relatively easy to include a finite temperature effect, as we demonstrate
below.

We consider a magnetized plasma with a background magnetic field $\mathbf{B} =
B_0 \mathbf{e}_x$. The density, and temperature are assumed to be constant. On
average (i.e, in the absence of transverse perturbation), the plasma is at
rest in the laboratory frame, so that the flow velocity along the magnetic field
is always zero $u_x = 0$. The homogeneity in density implies that the
longitudinal electric field is also zero $E_x = 0$. Now it is convenient to
introduce the following definitions for transverse perturbations:
\begin{align}
 B_{\perp} = B_y - i B_z, \quad
 E_{\perp} = E_y - i E_z, \quad
 u_{s,\perp} = u_{s,y} - i u_{s,z}.
\end{align}
Because there is no longitudinal perturbation, only the transverse components of
the equation of motion are non-trivial:
\begin{align}
 \frac{\partial}{\partial t} \left(\rho_s h_s \gamma_s u_{s,y}\right) =
 \mu_s \rho_s \gamma_s
 \left( E_y + \frac{u_{s,z}}{\gamma_s c} B_0 \right),
 \\
 \frac{\partial}{\partial t} \left(\rho_s h_s \gamma_s u_{s,z}\right) =
 \mu_s \rho_s \gamma_s
 \left( E_z - \frac{u_{s,y}}{\gamma_s c} B_0 \right),
\end{align}
where $h_s$ is the specific enthalpy defined by Eq.~(\ref{eq:enthalpy}),
representing an inertia increase due to a relativistically hot temperature. In
the absence of longitudinal perturbation, $h_s$ becomes just a constant,
indicating that the effect of finite temperature may be included by simply
replacing the mass $m_s$ of a cold fluid by $m_s h_s$.

Notice that, for circularly polarized modes, the Lorentz factor of the fluid
due to particle quiver motion is constant. Therefore, it can be put outside of
the temporal derivative, yielding
\begin{align}
 \frac{\partial}{\partial t} u_{s,\perp} =
 \frac{\mu_s}{h_s}
 E_{\perp} + i \frac{\bar{\Omega}_{c,s}}{\gamma_s} u_{s,\perp},
 \label{eq:cpw_eom}
\end{align}
where $\bar{\Omega}_{c,s} = \mu_s B_0 / h_s c$ is the effective cyclotron
frequency for particle species $s$.

Now we consider a monochromatic wave solution of the form $u_{s,\perp} =
\tilde{u}_{s,\perp} \exp \left(i k x - i \omega t \right)$. With this
definition, a solution in the domain $\omega > 0, k > 0$ corresponds to a
right-hand circularly polarized mode propagating in the positive $x$
direction. Similarly, we also define $\tilde{E}_{\perp}, \tilde{B}_{\perp}$ as
the wave amplitude. Eq.~(\ref{eq:cpw_eom}) then becomes
\begin{align}
 \tilde{u}_{s,\perp} =
 \frac{i}{\omega + \bar{\Omega}_{c,s}/\gamma_{s}}
 \frac{\mu_s}{h_s} \tilde{E}_{\perp},
 \label{eq:cpw_uperp}
\end{align}
whereas we have from Maxwell's equations
\begin{align}
 \left(1 - \frac{k^2 c^2}{\omega^2}\right) \tilde{E}_{\perp} &=
 - \frac{i}{\omega} \sum \mu_s \rho_s \tilde{u}_{s,\perp},
 \label{eq:cpw_ampare}\\
 \tilde{E}_{\perp} &= i \frac{\omega}{k c} \tilde{B}_{\perp}.
 \label{eq:cpw_faraday}
\end{align}
Combining Eqs.~(\ref{eq:cpw_uperp}) and (\ref{eq:cpw_ampare}), one obtains the
dispersion relation:
\begin{align}
 \omega^2 - k^2 c^2 - \sum
 \gamma_s \bar{\omega}_{p,s}^2
 \frac{\omega}{\gamma_s \omega + \bar{\Omega}_{c,s}} = 0,
 \label{eq:cpw_dispersion}
\end{align}
where $\bar{\omega}_{p,s}^2 = \mu_s^2 \rho_s / h_s$ is the effective proper
plasma frequency. Notice that, however, the above dispersion relation includes
the Lorentz factors $\gamma_s$ for each fluid, which are yet unknown at this
point.

To obtain the Lorentz factor, one may eliminate the electric field from
Eqs.~(\ref{eq:cpw_uperp}) and (\ref{eq:cpw_faraday}), yielding
\begin{align}
 \frac{\tilde{u}_{s,\perp}}{c} = - \gamma_s
 \frac{\bar{\Omega}_{c,s}}{\gamma_s \omega + \bar{\Omega}_{c,s}}
 \frac{\omega}{k c} \frac{\tilde{B}_{\perp}}{B_0}.
\end{align}
Squaring the equation and using $\gamma_s^2 = 1 + \tilde{u}_{\perp}^2/c^2$, we
get the following equation:
\begin{align}
 \gamma_s^4 +
 2 \frac{\bar{\Omega}_{c,s}}{\omega} \gamma_s^3 +
 \left( \frac{\bar{\Omega}_{c,s}^2}{\omega^2}
 \left(1 - \frac{\omega^2}{k^2 c^2} \xi^2 \right) - 1 \right) \gamma_s^2 -
 2 \frac{\bar{\Omega}_{c,s}}{\omega} \gamma_s -
 \frac{\bar{\Omega}_{c,s}^2}{\omega^2} = 0,
 \label{eq:cpw_gamma}
\end{align}
where the wave magnetic field amplitude normalized to the background is
denoted by $\xi = \tilde{B}_{\perp}/B_0$,

In summary, a finite amplitude circularly polarized wave solution for a given
set of $(k, \xi)$ is obtained by iteratively searching for a solution
$(\omega, \gamma_{p}, \gamma_{e})$ that simultaneously satisfies the three
equations: Eq.(\ref{eq:cpw_dispersion}) and (\ref{eq:cpw_gamma}) for each
fluid. In practice, one may adopt a simplified numerical procedure as
explained below. Given an initial guess ($\omega, \gamma_p, \gamma_e$), we
first solve Eq.~(\ref{eq:cpw_gamma}) independently for a given $\omega$. Then,
the dispersion relation (\ref{eq:cpw_dispersion}) is solved for given
$(\gamma_p, \gamma_e)$. The solution is obtained by iterating the whole
procedure. We find that even with this simplified iteration procedure, the
solution converges rather quickly.

For a fixed $k > 0$, the dispersion relation has two solutions in the region
$\omega > 0$; a subluminal $(\omega/k < c)$ \Alfven mode, and a superluminal
$(\omega/k > c)$ electromagnetic mode. Therefore, by starting from a small
$\omega$ as an initial guess, the solution will converge to the subluminal
mode. Conversely, a larger initial guess of $\omega$ will converge to the
superluminal mode. A left-hand circularly polarized wave solution can also be
obtained as a negative frequency root $\omega < 0$ of the same equation.

Notice that because the Lorentz factors of the two fluids are different, the
proper number densities must also be different to satisfy the charge
neutrality condition
\begin{align}
 \sum \mu_s \rho_s \gamma_s = \sum e_s n_s \gamma_s = 0.
\end{align}
Namely, it is the lab frame number density that must be equal between the
species. This condition is indeed necessary so that the dispersion relation
reduces to the \Alfven wave of RMHD in the low frequency limit $\omega \ll
\bar{\Omega}_{c,s}/\gamma_s$.

It is also worth noting that, for a given amplitude $\xi$, there exists a
critical wavenumber beyond which the \Alfven mode does not exist
\citep{2004physics..10203H}. This occurs due to increased particle inertia by
relativistic quiver motion, which decreases the effective cyclotron
frequency. For numerical benchmark problems, this does not pose any difficulty
by using a sufficiently small amplitude and/or wavenumber.

\newpage
\bibliographystyle{apj}
\bibliography{reference}
%\input{ms-arxiv.bbl}

%%% figures for manuscript style
\ifemulateapj
\else
\FigureOne
\FigureTwo
\FigureThree
\FigureFour
\FigureFive
\FigureSix
\FigureSeven
\FigureEight
\FigureNine
\FigureTen
\FigureEleven
\FigureTwelve
\fi
\end{document}